\documentclass[useAMS,usenatbib]{mn2e}
\pdfoutput=1
\usepackage{amsmath}
\usepackage{amsfonts}
\usepackage{amssymb}
\usepackage{bm}
\usepackage{booktabs}
\usepackage{natbib}
\usepackage{graphicx}
\usepackage{pgf}
\usepackage{ifthen}
\usepackage{tikz}
\usepackage{tikz-3dplot}
\usepackage[colorlinks=true,backref=page,bookmarksdepth=2]{hyperref}

\title[The three-dimensional problem of two fixed centres]
{
A complete and explicit solution to the three-dimensional problem of two fixed centres
}

\author[Francesco Biscani and Dario Izzo]{Francesco Biscani$^1$\thanks{E-mail: bluescarni@gmail.com} and Dario Izzo$^2$\\
$^{1}$Max Planck Institute for Astronomy,\\
$\hphantom{^{2}}$K\"{o}nigstuhl 17, D-69117 Heidelberg, Germany\\
$^{2}$ESA -- Advanced Concepts Team, European Space Research Technology Centre (ESTEC),\\
$\hphantom{^{2}}$Keplerlaan 1, Postbus 299, 2200 AG Noordwijk The Netherlands
}

\begin{document}
\date{\today}

\pagerange{\pageref{firstpage}--\pageref{lastpage}}

\maketitle

\label{firstpage}

\begin{abstract}
We present for the first time an explicit, complete and closed-form solution to the three-dimensional problem of two fixed centres, based on Weierstrass
elliptic and related functions. With respect to previous treatments of the problem, our solution is exact, valid for all initial conditions
and physical parameters of the system (including unbounded orbits and repulsive forces), and expressed via a unique set of formul\ae{}.
Various properties of the three-dimensional problem of two fixed centres are investigated and analysed, with a particular emphasis on quasi-periodic and periodic orbits, regions of motion
and equilibrium points.
\end{abstract}

\begin{keywords}
Celestial mechanics - Gravitation
\end{keywords}
\section{Introduction}
The problem of two fixed centres is the dynamical system consisting of two fixed bodies exerting inverse-square forces on a test particle.
It is also known as \emph{Euler's three-body problem} (E3BP), the \emph{Euler-Jacobi problem}, and the \emph{two-centre Kepler problem}, and it
is the simplest three-body problem of physical interest.
The solvability of the E3BP was first established by Euler in the 1760s, who demonstrated the existence of integrals of motion involving inverse square roots of quartic
expressions. Later on, the E3BP attracted the attention of Legendre, Lagrange and Jacobi, who recognized that the solution of the E3BP can be expressed in terms
of elliptic functions and integrals.

Throughout the twentieth century, mathematicians and astronomers frequently returned to the E3BP. \citet{darboux_sur_1901} showed
how the E3BP can be generalised with the addition of complex masses at complex distances, while retaining separability in the elliptic coordinate system.
\citet{charlier_mechanik_1902} used the E3BP as the starting point for a discussion of the restricted three-body problem in his treatise on celestial mechanics.
\citet{hiltebeitel_problem_1911} introduced further generalisations of the E3BP involving linear and inverse-cube forces.

In the early days of quantum mechanics, the E3BP served as a model of the hydrogen molecule ion $\textnormal{H}_2^+$ \citep{pauli_uber_1922}. The advent of the space age
brought a renewed interest in the E3BP, after it was established that the it could be used as an approximation of the potential of rotationally-symmetric rigid bodies
\citep{vinti_new_1959,demin_orbits_1961,aksenov_general_1962,deprit_probleme_1962,aksenov_generalized_1963,alfriend_extended_1977}.
Even recently, the E3BP has been the subject of ongoing research, both in its classical formulation but also in connection with quantum mechanics and general relativity
\citep{cordani_kepler_2003,varvoglis_two_2004,waalkens_problem_2004,coelho_relativistic_2009}. The work of \citet{omathuna_integrable_2008} is particularly noteworthy, devoting four chapters to the E3BP.

From a purely mathematical point of view, the interest in the E3BP arises from the fact that it belongs to the very restrictive class of Liouville-integrable dynamical systems
\citep{arnold_mathematical_1989}. From a physical and astronomical point of view, the E3BP is noteworthy for at least two reasons:
\begin{itemize}
 \item as a (completely solvable) stepping-stone between the two-body problem and the three-body problem. In particular, the planar E3BP is analogous to a circular restricted three-body problem without centrifugal force;
 \item as an approximation of the potential of a rotationally-symmetric rigid body.
\end{itemize}

Despite more than two centuries of research, a full and explicit solution to the E3BP has so far proven elusive. As pointed out by \citet{omathuna_integrable_2008}, most authors explain
that the solution involves elliptic functions and integrals, but they stop short of explicitly computing such solution or even sketching out how it is to be attained. This is
particularly true for the three-dimensional E3BP, which is typically treated as an extension of the bidimensional case. Even in the very thorough exposition of \citet{omathuna_integrable_2008},
the solution of the three-dimensional E3BP is limited to the case of negative energy and it is not explicit in the third coordinate and in the time-angle relation.

The aim of this paper is to present for the first time a full explicit solution to the three-dimensional E3BP in terms of Weierstrass elliptic and related functions.
Our solution is expressed via unique formul\ae{}
valid for any set of physical parameters and initial conditions of the three-dimensional case. The use of the Weierstrassian formalism allows us to provide a compact
formulation for the solution and to avoid the fragmentation usually associated to the use of Jacobi elliptic functions.

We need to point out that our focus
is the three-dimensional case, and although the method presented here might also be suitable for the solution of the bidimensional case, we will not consider
planar motion: our solution to the three-dimensional case stands alone and it is not formulated as a generalisation of the bidimensional case.
We also need to make clear that the main objective of this work is to detail the derivation of our solution and to present its general mathematical properties.
The application to problems of astronomical and physical interest is left for subsequent publications.

The paper is structured as follows:
in Section \ref{sec:formulation}, we formulate the problem and we identify the integrals of motion in the Hamiltonian formalism;
in Section \ref{sec:integration}, we solve the equations of motion and the time equation in elliptic-cylindrical coordinates;
Section \ref{sec:analysis} is dedicated to the analysis of the solution, which focuses on periodic and quasi-periodic orbits, regions of motion,
 equilibrium points and displaced circular orbits;
Section \ref{sec:conclusions} is dedicated to the conclusions and to possible future extensions of the work presented here.

\section{Formulation of the problem}
\label{sec:formulation}
\begin{figure}
\begin{center}
\tikzset{
  center/.style={
    circle, inner sep=0pt, 
    minimum size=2mm, fill=gray, draw=black
 },
  particle/.style={
    circle, inner sep=0pt, 
    minimum size=.8mm, fill=white, draw=black
 },
}
\tdplotsetmaincoords{60}{120}
\begin{tikzpicture}
	[scale=4,
		tdplot_main_coords,
		axis/.style={->,black,thick},
		axis dash/.style={black,thick,dashed},
		force/.style={->,black},
		vector guide/.style={dashed,draw=gray},
		vector guide arrow/.style={->,draw=gray},
		arrow/.style={<->,very thin,draw=gray}]

	\coordinate (O) at (0,0,0);

	\draw[axis] (0,0,0) -- (1,0,0) node[anchor=north east]{$x$};
	\draw[axis] (0,0,0) -- (0,1,0) node[anchor=north west]{$y$};
	\draw[axis] (0,0,0) -- (0,0,.75) node[anchor=south]{$z$};
	\draw[axis dash] (0,0,-.75) -- (0,0,0) {};
	
	\node[center] (mu1) at (0,0,.5) {};
	\node[anchor=east] () at (mu1) {$\mu_1\,$};
	\node[center] (mu2) at (0,0,-.5) {};
	\node[anchor=east] () at (mu2) {$\mu_2\,$};

	\coordinate (P) at (.5,1,.9);
	\coordinate (Pxy) at (.5,1,0.);

	\node[particle] (particle) at (P) {};

	\draw[vector guide arrow] (O) -- (Pxy) node[sloped,midway,below] {\rotatebox[origin=c]{10}{$\textnormal{\hphantom{a}}\rho$}};
	\draw[vector guide] (Pxy) -- (particle);
	\draw[vector guide arrow] (O) -- (particle) node[sloped,midway,below] {$\bm{r}$};

	\node[anchor=west] () at (particle) {$P$};

	\draw[force] (particle) -- (mu1) node[sloped,midway,below]{$\bm{F}_1$};
	\draw[force] (particle) -- (mu2) node[sloped,midway,below]{$\textnormal{\hphantom{aaa}}\bm{F}_2$};
	
	\draw[arrow] (.03,-.03,0) -- (.03,-.03,.47) node[sloped,midway,above] {$a$};
	\draw[arrow] (.03,-.03,-.47) -- (.03,-.03,0) node[sloped,midway,above] {$a$};
	
	\tdplotdrawarc[vector guide]{(O)}{.3}{0}{60}{anchor=south}{$\textnormal{\hphantom{aaa}}\phi$}
\end{tikzpicture}
\end{center}
\caption{Setup of the three-dimensional E3BP. Two fixed centres $\mu_1$ and $\mu_2$ are located on the positive and negative $z$ axis at a distance
$a$ from the origin of an inertial frame of reference. $\mu_1$ and $\mu_2$ exert inverse-square forces $\bm{F}_1$ and $\bm{F}_2$ (attractive or repulsive) on a test particle $P$ located at the position $\bm{r}$.
The cylindrical radius $\rho$ is the magnitude of the projection of $\bm{r}$ on the $xy$ plane. $\phi$ is the azimuthal angle.\label{fig:setup}}
\end{figure}
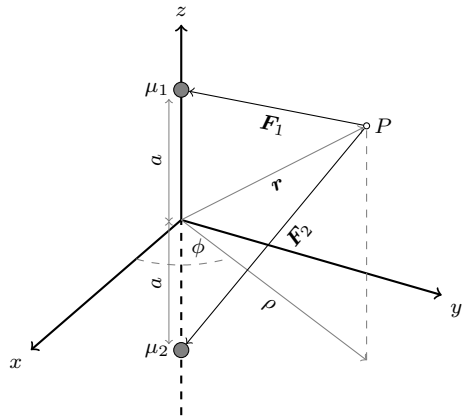

The setup of the three-dimensional E3BP is sketched in Figure \ref{fig:setup}. Without loss of generality, we can position the two centres on the positive and negative $z$ axis at a distance
$a$ from the origin of an inertial reference frame. This choice naturally allows to take immediate advantage of the cylindrical symmetry of the problem, as it will be shown
shortly. The Lagrangian of a test particle moving in the potential generated by the two fixed centres reads:
\begin{multline}
\mathcal{L}\left( \bm{r};\Dot{\bm{r}} \right) = \frac{\Dot{\bm{r}}^2}{2} + \frac{\mu_1}{\sqrt{x^2+y^2+\left( z - a\right)^2}} \\
+ \frac{\mu_2}{\sqrt{x^2+y^2+\left( z + a\right)^2}},\label{eq:lagr00}
\end{multline}
where $\bm{r}=\left(x,y,z\right)$ is the vector of cartesian coordinates of the test particle, $\Dot{\bm{r}}$ its time derivative, and $\mu_{1,2}$ represent the strength
of the two centres of attraction. $\mu_{1,2}$ can either be either positive or negative, resulting respectively in attractive or repulsive forces on the test particle.
In the gravitational E3BP, $\mu_{1,2}$ are gravitational parameters:
\begin{equation}
\mu_{1,2} = \mathcal{G}M_{1,2},
\end{equation}
where $M_{1,2}$ are the masses of the two centres and $\mathcal{G}$ is the gravitational constant. Although our solution is valid for both negative and positive
$\mu_{1,2}$, in the analysis of the results we will focus on the gravitational E3BP, as it is arguably the most useful version of the E3BP in an astronomical context.

Following \citet{waalkens_problem_2004}, the first step of the separation of the problem is the introduction of cylindrical coordinates:
\begin{align}
x & = \rho\cos\phi, & \rho & = \sqrt{x^2 + y^2}, \\
y & = \rho\sin\phi, & \phi & = \arctan \left( y,x \right),
\end{align}
for the coordinates and
\begin{align}
\Dot{x} & = \Dot{\rho}\cos\phi - \rho\Dot{\phi}\sin\phi, & \Dot{\rho} & = \frac{x\Dot{x}+y\Dot{y}}{\sqrt{x^2 + y^2}}, \\
\Dot{y} & = \Dot{\rho}\sin\phi + \rho\Dot{\phi}\cos\phi, & \dot{\phi} & =\frac{\dot{y}x-\dot{x}y}{x^{2}+y^{2}},
\end{align}
for the velocities, where $\arctan\left(\right)$ is the two-argument inverse tangent function.
The Lagrangian in cylindrical coordinates reads:
\begin{multline}
\mathcal{L}\left( \rho,\phi,z;\Dot{\rho},\Dot{\phi},\Dot{z} \right) = \frac{\Dot{\rho}^2 + \rho^2\Dot{\phi}^2+\Dot{z}^2}{2} \\
+ \frac{\mu_1}{\sqrt{\rho^2+\left( z - a\right)^2}}
+ \frac{\mu_2}{\sqrt{\rho^2+\left( z + a\right)^2}}.
\end{multline}
As a result of the rotational symmetry of the system, the $\phi$ coordinate is now cyclic and the Lagrangian is reduced
to two degrees of freedom (in the variables $\rho$ and $z$).

We now proceed to introduce elliptical coordinates on the $\left( \rho, z \right)$ plane via the transformation
\begin{align}
\rho & = a\sqrt{\left( \xi^2 - 1 \right)\left( 1 - \eta^2 \right)},\label{eq:rho_ell}\\
z & = a\xi\eta,\label{eq:z_ell}
\end{align}
its inverse
\begin{align}
\xi & = \frac{\sqrt{\rho^2+\left( z + a\right)^2} + \sqrt{\rho^2+\left( z - a\right)^2}}{2a},\label{eq:xidef} \\
\eta & = \frac{\sqrt{\rho^2+\left( z + a\right)^2}-\sqrt{\rho^2+\left( z - a\right)^2}}{2a},\label{eq:etadef}
\end{align}
and the corresponding transformations for the velocities,
\begin{align}
\Dot{\rho} & = a\frac{\Dot{\xi}\xi\left( 1 - \eta^2 \right) - \Dot{\eta}\eta\left( \xi^2 - 1 \right)}{\sqrt{\left( \xi^2 - 1 \right) \left( 1 - \eta^2 \right)}}, \\
\Dot{z} & = a\left( \Dot{\xi}\eta + \xi\Dot{\eta} \right),
\end{align}
and
\begin{align}
\Dot{\xi} & = \frac{1}{2a}\left[ \frac{\Dot{\rho}\rho+\Dot{z}\left( z + a \right)}{\sqrt{\rho^2+\left( z + a \right)^2}} +
\frac{\Dot{\rho}\rho+\Dot{z}\left( z - a \right)}{\sqrt{\rho^2+\left( z - a \right)^2}}\right],\\
\Dot{\eta} & = \frac{1}{2a}\left[ \frac{\Dot{\rho}\rho+\Dot{z}\left( z + a \right)}{\sqrt{\rho^2+\left( z + a \right)^2}} -
\frac{\Dot{\rho}\rho+\Dot{z}\left( z - a \right)}{\sqrt{\rho^2+\left( z - a \right)^2}}\right].
\end{align}
The right-hand side of eq. \eqref{eq:xidef} represents the sum of the distances of the point
$\left( \rho,z\right)$ from the two centres, normalised by $2a$. Similarly, the right-hand side of eq. \eqref{eq:etadef} represents the
difference of the distances of the point
$\left( \rho,z\right)$ from the two centres, normalised by $2a$. It thus follows that the lines of constant $\xi$ and $\eta$ in the $\left( \rho,z\right)$
plane are ellipses and branches of hyperbol\ae{} whose foci are coincident with the masses. The domain of $\xi$ is $\left[1,\infty\right)$, the domain of
$\eta$ is $\left[-1,1\right]$.

The Lagrangian in elliptic-cylindrical coordinates reads:
\begin{multline}
\mathcal{L}\left( \xi,\eta, \phi;\Dot{\xi},\Dot{\eta},\Dot{\phi} \right) = \frac{a^2}{2} \left[ \frac{\Dot{\xi}^2\xi^2\left(1-\eta^2\right)}{\xi^2 - 1} +
\frac{\Dot{\eta}^2\eta^2\left(\xi^2 - 1\right)}{1-\eta^2}\right.\\
\left.  \vphantom{\frac{\Dot{\xi}^2\xi^2\left(1-\eta^2\right)}{\xi^2 - 1}}
+ \Dot{\phi}^2\left( \xi^2 - 1\right)\left(1 - \eta^2\right) + \Dot{\xi}^2\eta^2 + \xi^2\Dot{\eta}^2\right]\\
+\frac{\mu_1}{a\left(\xi - \eta\right)} + \frac{\mu_2}{a\left(\xi + \eta\right)}.
\end{multline}
Switching now to the Hamiltonian formulation of the problem with a Legendre transformation, we introduce the generalised momenta via the relations
\begin{align}
p_\xi & = \frac{\partial \mathcal{L}}{\partial \Dot{\xi}} = a^2\Dot{\xi}\left(\frac{\xi^2-\eta^2}{\xi^2-1}\right),\\
p_\eta &  = \frac{\partial \mathcal{L}}{\partial \Dot{\eta}} = a^2\Dot{\eta}\left(\frac{\xi^2-\eta^2}{1-\eta^2}\right),\\
p_\phi &  = \frac{\partial \mathcal{L}}{\partial \Dot{\phi}} = a^2\Dot{\phi}\left( \xi^2 - 1\right)\left(1 - \eta^2\right),
\end{align}
and the inverse
\begin{align}
\Dot{\xi} & = \frac{p_\xi \left( \xi^2 - 1 \right) }{a^2\left( \xi^2-\eta^2 \right)},\\
\Dot{\eta} & = \frac{p_\eta \left( 1 - \eta^2\right) }{a^2\left( \xi^2-\eta^2 \right)},\\
\Dot{\phi} & = \frac{p_\phi}{a^2\left( \xi^2 - 1 \right) \left( 1 - \eta^2\right)}.
\end{align}

The Hamiltonian then reads:
\begin{multline}
\mathcal{H}\left( p_\xi,p_\eta, p_\phi;\xi,\eta, \phi \right) = \\
\frac{1}{2a^2\left( \xi^2 - \eta^2 \right)}\left[p_\xi^2\left(\xi^2 -1 \right) + p_\eta^2\left(1-\eta^2\right)\right]
\\ +\frac{p_\phi^2}{2a^2\left(\xi^2 - 1\right)\left(1-\eta^2\right)} - \frac{\mu_1}{a\left(\xi - \eta\right)} - \frac{\mu_2}{a\left(\xi + \eta\right)}.
\label{eq:t_Ham}
\end{multline}
$p_\phi$ is a constant of motion equal to the $z$ component of the angular momentum per unit mass of the test particle.
When $p_\phi$ is zero, the trajectory of the particle crosses the $z$ axis and it remains constrained to a plane (that is,
the problem is reduced to the planar case). As explained earlier, we will not
consider here this special case, and we always assume that $p_\phi$ is nonzero (or, equivalently, $\xi \neq 1$ and $\eta \neq \pm 1$).

We proceed now to the introduction of the new Hamiltonian $\mathcal{H}_\tau$, obtained through the following
Poincar\'{e} time transformation \citep{siegel_lectures_1971,carinena_time_1988,saha_interpreting_2009}:
\begin{equation}
\mathcal{H}_\tau = \left(\mathcal{H} - h\right) \left( \xi^2 - \eta^2 \right),
\end{equation}
where $h$ is the energy constant of the system (i.e., the numerical value of $\mathcal{H}$ after the substitution
of the initial conditions into \eqref{eq:t_Ham}).
The Hamiltonian $\mathcal{H}_\tau$ generates the equations of motion with respect to the new independent
variable $\tau$ (often called a \emph{fictitious} time), connected to the real time $t$ through the differential relation
\begin{equation}
dt=\left(\xi^{2}-\eta^{2}\right)d\tau. \label{eq:fic_time}
\end{equation}
The Poincar\'{e} time transform can be seen as the Hamiltonian analogue of a Sundman transformation \citep{sundman_memoire_1912}:
it regularises the problem by slowing time down when the particle is close to either fixed centre (note how in correspondence of the two centres
$\xi^{2}-\eta^{2}$ is zero).

The equations of motion in fictitious time read:
\begin{align}
\frac{d\xi}{d\tau} & = \frac{\partial \mathcal{H}_\tau}{\partial p_\xi}, & \frac{dp_\xi}{d\tau} & = -\frac{\partial \mathcal{H}_\tau}{\partial \xi}, \label{eq:xi_tau} \\
\frac{d\eta}{d\tau} & = \frac{\partial \mathcal{H}_\tau}{\partial p_\eta}, & \frac{dp_\eta}{d\tau} & = -\frac{\partial \mathcal{H}_\tau}{\partial \eta}, \label{eq:eta_tau}\\
\frac{d\phi}{d\tau} & = \frac{\partial \mathcal{H}_\tau}{\partial p_\phi}, & \frac{dp_\phi}{d\tau} & = -\frac{\partial \mathcal{H}_\tau}{\partial \phi}, \label{eq:phi_tau}
\end{align}
 while the complete expression for $\mathcal{H}_\tau$ is
\begin{multline}
\mathcal{H}_\tau \left( p_\xi,p_\eta, p_\phi;\xi,\eta, \phi \right) = \\
- \xi^{2} h - \frac{\xi}{a} \left(\mu_{1} + \mu_{2}\right) + \frac{p_{\phi}^{2}}{2 a^{2} \left(\xi^{2} - 1\right)} + \frac{p_{\xi}^{2}}{2 a^{2}} \left(\xi^{2} - 1\right)\\
+\eta^{2} h - \frac{\eta}{a} \left(\mu_{1} - \mu_{2}\right) + \frac{p_{\phi}^{2}}{2 a^{2} \left(1 - \eta^{2}\right)} + \frac{p_{\eta}^{2}}{2 a^{2}} \left(1 - \eta^{2}\right).
\end{multline}
The Hamiltonian $\mathcal{H}_\tau$ has thus been separated into the two constants of motion
\begin{align}
h_\xi & = - \xi^{2} h - \frac{\xi}{a} \left(\mu_{1} + \mu_{2}\right) + \frac{p_{\phi}^{2}}{2 a^{2} \left(\xi^{2} - 1\right)} + \frac{p_{\xi}^{2}}{2 a^{2}} \left(\xi^{2} - 1\right), \label{eq:h_xi}\\
h_\eta & = \eta^{2} h - \frac{\eta}{a} \left(\mu_{1} - \mu_{2}\right) + \frac{p_{\phi}^{2}}{2 a^{2} \left(1 - \eta^{2}\right)} + \frac{p_{\eta}^{2}}{2 a^{2}} \left(1 - \eta^{2}\right). \label{eq:h_eta}
\end{align}
We can now move on to the solution of the equations of motion.

\section{Integration of the equations of motion}
\label{sec:integration}
We proceed now to the explicit integration of the equations of motion \eqref{eq:xi_tau}--\eqref{eq:phi_tau} and of the time equation
\eqref{eq:fic_time}. We detail initially the solution for the $\xi$ coordinate;
the solution for the $\eta$ coordinate if formally identical, while the solutions for the $\phi$ coordinate and for the time equation require a different procedure.
The solutions for the momenta $p_\xi$ and $p_\eta$ will be easily computed via the solutions for $\xi$ and $\eta$.
In this section
we will focus on the mathematical details of the solution. Section \ref{sec:analysis} will be dedicated to the analysis of our results.
\subsection{Solution for $\xi$}
\label{subsec:xi_sol}
The equation of motion for $\xi$, eq. \eqref{eq:xi_tau}, reads:
\begin{equation}
\frac{d\xi}{d\tau} = \frac{\xi^{2} p_{\xi}}{a^{2}} - \frac{p_{\xi}}{a^{2}}.\label{eq:xi_mot}
\end{equation}
We can use the constant of motion $h_\xi$ to express $p_\xi$ as a function of $\xi$, via the inversion of eq. \eqref{eq:h_xi}:
\begin{equation}
p_\xi = \pm \frac{1}{\xi^2 - 1}\sqrt{f_\xi\left( \xi \right)},
\end{equation}
where
\begin{multline}
f_\xi\left( \xi \right) = 2 \xi^{4} a^{2} h + \xi^{3} \left(2 \mu_{1} a + 2 \mu_{2} a\right) \\
+ \xi^{2} \left(- 2 a^{2} h + 2 a^{2} h_{\xi}\right) + \xi \left(- 2 \mu_{1} a - 2 \mu_{2} a\right) \\
- 2 a^{2} h_{\xi} - p_{\phi}^{2}\label{eq:xi_poly}
\end{multline}
is a polynomial of degree 4 in $\xi$. The equation of motion for $\xi$ then reads:
\begin{equation}
\frac{d\xi}{d\tau} = \pm \frac{\sqrt{f_\xi\left( \xi \right)}}{a^2}.\label{eq:dxi_dtau_00}
\end{equation}

\begin{figure*}
\includegraphics[width=1\textwidth]{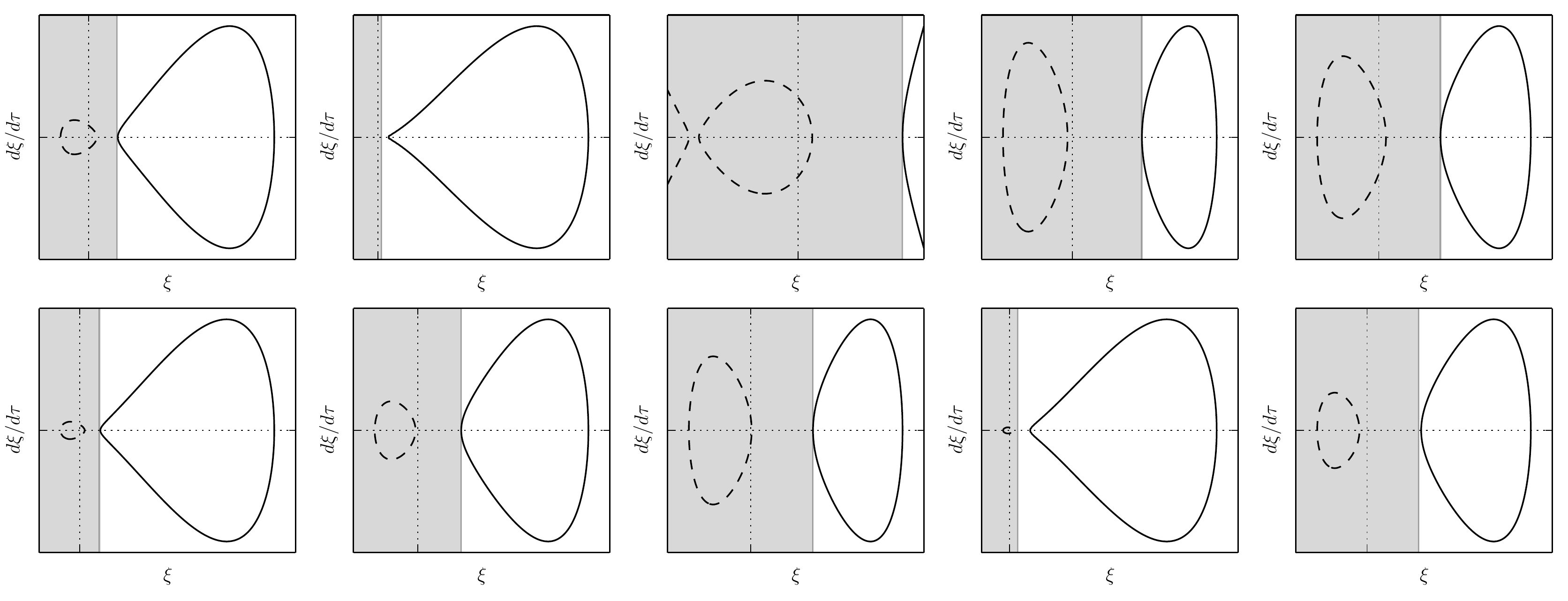}

\caption{Phase portraits for the $\xi$ variable in ten randomly-generated cases. The shaded region corresponding to $\xi < 1$ is the domain
in which the $\xi$ variable does not represent any real physical coordinate. The trajectories in the $\left( \xi, d\xi/d\tau \right)$ plane
are elliptic curves. All the portraits refer to bounded trajectories, apart from the third panel in the first row.\label{fig:xi_poly}}
\end{figure*}

Before proceeding to the integration and inversion of this equation, we need to highlight a useful property of
$f_\xi\left( \xi \right)$. It can be checked by direct substitution that
\begin{equation}
f_\xi\left( 1 \right) = -p_\phi^2 < 0,
\end{equation}
that is, the value of the polynomial is always negative for $\xi = 1$. On the other hand, for any choice of initial conditions,
$f_\xi\left( \xi \right)$ must assume positive values in a subrange of $\xi > 1$, otherwise the radicand in eq. \eqref{eq:dxi_dtau_00} would be negative.
Consequently, the polynomial $f_\xi\left( \xi \right)$ must have at least one real root in the range $\left(1,+\infty\right)$ (i.e., in the domain of
interest for the variable $\xi$). Figure \ref{fig:xi_poly} visualises the phase portraits for $\xi$ in a few randomly-selected cases. It should be noted
how a necessary condition for unbounded motion is a positive energy constant $h$: according to eqs. \eqref{eq:xi_poly} and \eqref{eq:dxi_dtau_00}, if $h$ is negative
the radicand on the right-hand side of eq. \eqref{eq:dxi_dtau_00} assumes negative values for $\xi\to\infty$. Thus, for negative $h$ the phase portrait for the $\xi$ variable
must be a closed curve.

The integration by quadrature of eq. \eqref{eq:dxi_dtau_00} reads:
\begin{equation}
\int_{\xi_0}^\xi\frac{ds}{\sqrt{f_\xi\left( s \right)}}=\pm\frac{1}{a^2}\int_{0}^\tau ds,\label{eq:xi_quad_00}
\end{equation}
where $s$ is a dummy integration variable, $\xi_0$ is an arbitrary initial value for $\xi$ and, without loss of generality,
we have set the initial fictitious time to $0$. The $\pm$ sign in front of the integral on the right-hand side can be chosen
as the sign of the initial value of $d\xi/d\tau$. With this convention, the right-hand side of eq. \eqref{eq:xi_quad_00} represents the time
needed by the system to evolve from $\xi_0$ to $\xi$ along a trajectory in which the sign of $d\xi/d\tau$ is constant.

The left-hand side of eq. \eqref{eq:xi_quad_00} is an elliptic integral which can be inverted to yield $\xi\left( \tau \right)$
using a formula by Weierstrass, reported in \citet[\S 20.6]{whittaker_course_1927}.
The general formula is rather complicated, but it can be simplified considerably by choosing as the initial value for the integration
a root $\xi_r$ of the polynomial $f_\xi\left( \xi \right)$. With such a choice, eq. \eqref{eq:xi_quad_00} reads
\begin{equation}
\int_{\xi_r}^\xi\frac{ds}{\sqrt{f_\xi\left( s \right)}}=\pm\frac{1}{a^2}\int_{\tau_\xi}^\tau ds,\label{eq:xi_quad_01}
\end{equation}
where now $\tau_\xi$ is the fictitious \emph{time of root passage} for $\xi$. In analogy with the Kepler problem,
when $\xi_r$ is the smallest reachable root in the domain of interest $\tau_\xi$ can be considered as a kind of fictitious ``time of pericentre passage''
for the $\xi$ variable.
$\tau_\xi$ can be computed by solving the elliptic integral:
\begin{equation}
\tau_\xi = \pm a^2\int_{\xi_0}^{\xi_r}\frac{ds}{\sqrt{f_\xi\left( s \right)}}.\label{eq:tau_xi}
\end{equation}
As we have mentioned above, the existence of at least one root $\xi_r$ in the domain of interest for $\xi$
is guaranteed by the properties of the coefficients of $f_\xi\left( \xi \right)$. $\xi_r$ can be computed either numerically or algebraically (via the quartic
formula).

Equation \eqref{eq:xi_quad_01} can now be integrated and inverted to yield
\begin{equation}
\xi\left( \tau \right) = \xi_r + \frac{1}{4}\frac{f_\xi^\prime\left( \xi_r \right)}{\wp_\xi\left( \frac{\tau - \tau_\xi}{a^2} \right) - \frac{1}{24}f_\xi^{\prime\prime}\left( \xi_r \right)}.\label{eq:xi_sol}
\end{equation}
In this formula, the derivatives of $f_\xi\left( \xi \right)$ are to be taken with respect to $\xi$, and $\wp_\xi$ is a Weierstrass elliptic function defined in terms of the
two \emph{invariants}
\begin{align}
g_2 & = \mu_{1}^{2} a^{2} + 2 \mu_{1} \mu_{2} a^{2} + \mu_{2}^{2} a^{2} + \frac{a^{4} h^{2}}{3} - \frac{14 h}{3} a^{4} h_{\xi}\notag\\
& \quad + \frac{a^{4} h_{\xi}^{2}}{3} - 2 a^{2} h p_{\phi}^{2},\label{eq:g2_xi}\\
g_3 & = - \frac{\mu_{1}^{2} h}{3} a^{4} + \frac{\mu_{1}^{2} h_{\xi}}{3} a^{4} + \frac{\mu_{1}^{2} p_{\phi}^{2}}{4} a^{2} \notag\\
&\quad - \frac{2 \mu_{1}}{3} \mu_{2} a^{4} h + \frac{2 \mu_{1}}{3} \mu_{2} a^{4} h_{\xi} + \frac{\mu_{1} \mu_{2}}{2} a^{2} p_{\phi}^{2} \notag\\
&\quad - \frac{\mu_{2}^{2} h}{3} a^{4} + \frac{\mu_{2}^{2} h_{\xi}}{3} a^{4} + \frac{\mu_{2}^{2} p_{\phi}^{2}}{4} a^{2}\notag\\
&\quad + \frac{a^{6} h^{3}}{27} + \frac{11 h_{\xi}}{9} a^{6} h^{2} - \frac{11 h}{9} a^{6} h_{\xi}^{2} - \frac{a^{6} h_{\xi}^{3}}{27} \notag\\
&\quad + \frac{2 h^{2}}{3} a^{4} p_{\phi}^{2} - \frac{2 h}{3} a^{4} h_{\xi} p_{\phi}^{2},\label{eq:g3_xi}
\end{align}
computed from the coefficients of the polynomial $f_\xi\left( \xi \right)$ following \citet[\S 20.6]{whittaker_course_1927}.
Without giving a full account of the theory of the Weierstrassian functions
(for which we refer to standard textbooks such as \citet{whittaker_course_1927}, \citet{abramowitz_handbook_1964}, \citet{akhiezer_elements_1990}), we will recall here briefly a few fundamental notions about
the $\wp$ function. As commonly done, in the following we will suppress the verbose notation $\wp_\xi\left(\tau;g_2,g_3\right)$ in favour of just $\wp_\xi\left(\tau\right)$, with the
understanding that $\wp_\xi$ refers to a Weierstrass function defined in terms of the invariants \eqref{eq:g2_xi} and \eqref{eq:g3_xi}.

The elliptic function $\wp\left( z;g_2,g_3\right)$ is a doubly-periodic complex-valued function of a complex variable $z$ defined in terms of the two
invariants $g_2$ and $g_3$. The complex primitive half-periods
of $\wp$ can be related to the invariants via formul\ae{}
involving elliptic integrals and the roots $e_1$, $e_2$ and $e_3$ of the Weierstrass cubic
\begin{equation}
4t^3-g_2t-g_3 = 0 \label{eq:cubic_wp}
\end{equation}
\citep[e.g., see][\S 18.9]{abramowitz_handbook_1964}. The sign of the \emph{modular discriminant}
\begin{equation}
\Delta = g_2^3-27g_3^2
\end{equation}
determines the nature of the roots $e_1$, $e_2$ and $e_3$. In this specific case, the invariants are by definition real, $\tau$ is also real,
and $\wp_\xi\left(\tau\right)$ can thus be regarded
as a real-valued singly-periodic function on the real axis \citep[see][Chapter 18]{abramowitz_handbook_1964}. We refer to the real half-period of $\wp_\xi$ as $\omega_\xi$. It should be noted that,
according to eq. \eqref{eq:tau_xi}, if the real period of $\wp_\xi$ is $2\omega_\xi$, then the period
of $\xi\left(\tau\right)$ is $2a^2\omega_\xi$.

\begin{figure}
\includegraphics[width=.47\textwidth]{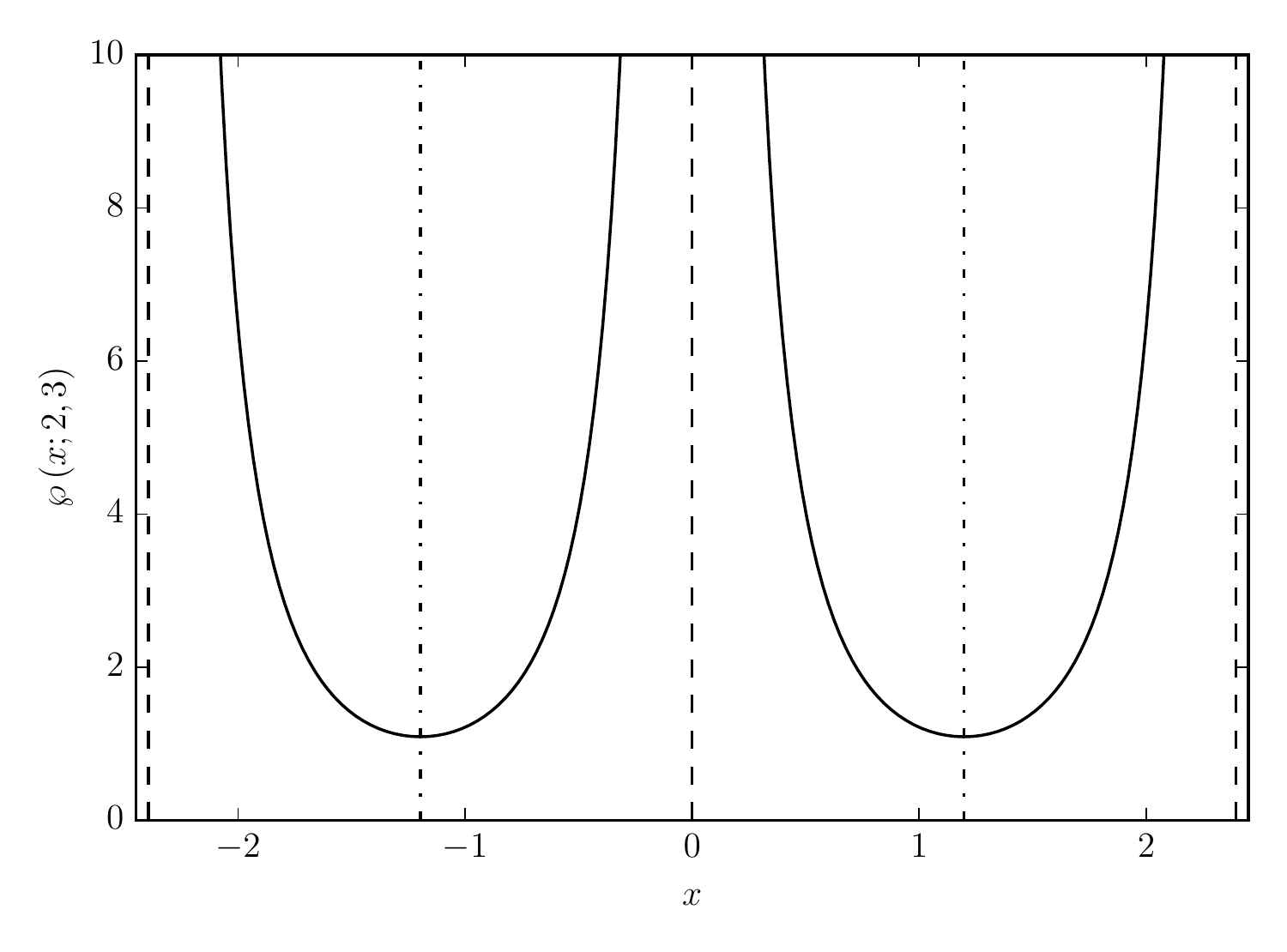}

\caption{Plot of $\wp\left(2,3;z\right)$ on the real axis over two real periods (solid line). In this specific case, the real period $2\omega$ is approximately
2.39. The vertical dashed lines represent asymptotes, close to which $\wp$ behaves like $1/z^2$ close to zero. The vertical dash-dotted lines in correspondence
of the real half-period $\omega$ cross $\wp$ at its absolute minimum value, and they are axes of symmetry for $\wp$ within the real period.\label{fig:wp}}
\end{figure}

$\wp$ is an even function, and it has second-order poles at each point of the period lattice. That is, $\wp_\xi$ satisfies the following properties
(where $k \in \mathbb{Z}$):
\begin{align}
\wp_\xi\left(\tau\right) & = \wp_\xi\left(\tau + 2k\omega_\xi\right),\\
\wp_\xi\left(\tau\right) & = \wp_\xi\left(-\tau\right),\\
\lim_{\tau \to 2k\omega_\xi}\wp_\xi\left(\tau\right) & = +\infty,
\end{align}
and in proximity of $\tau = 2k\omega_\xi$ the function $\wp_\xi\left(\tau\right)$ behaves, on the real axis, like $1/\tau^2$ in proximity of $\tau = 0$.
Figure \ref{fig:wp} illustrates graphically the behaviour of the Weierstrass elliptic function on the real axis.

We can also use the inverse Weierstrass elliptic function $\wp^{-1}$ to express the time of root passage via the inversion of eq. \eqref{eq:xi_sol} after setting $\tau = 0$:
\begin{equation}
\tau_\xi = a^2 \wp_\xi^{-1}\left[\frac{1}{24} f_\xi^{\prime\prime}\left( \xi_r \right) + \frac{f_\xi^\prime\left( \xi_r \right)}{4\left(\xi_0 - \xi_r\right)}\right].\label{eq:tau_xi_0}
\end{equation}
This is an alternative (but equivalent) formulation of the elliptic integral \eqref{eq:tau_xi}.
It must be noted that the function $\wp^{-1}$ has two solutions within the fundamental parallelogram of $\wp$, and thus two possible values for $\tau_\xi$ exist.
This ambiguity physically translates to the fact that each value assumed
by the $\xi$ coordinate is visited twice along a trajectory, once with a positive value for $d\xi/d\tau$ and once with a negative value\footnote{Another manifestation of this ambiguity
is the symmetry with respect to the horizontal axis of the phase portraits - see the examples in Figure \ref{fig:xi_poly}.}. The correct choice for $\tau_\xi$ is the value which,
when plugged into eq. \eqref{eq:pxi_sol} (which describes the evolution of $p_\xi$ in fictitious time), returns the initial value of $p_\xi$.

\subsection{Solution for $\eta$}
\begin{figure*}
\includegraphics[width=1\textwidth]{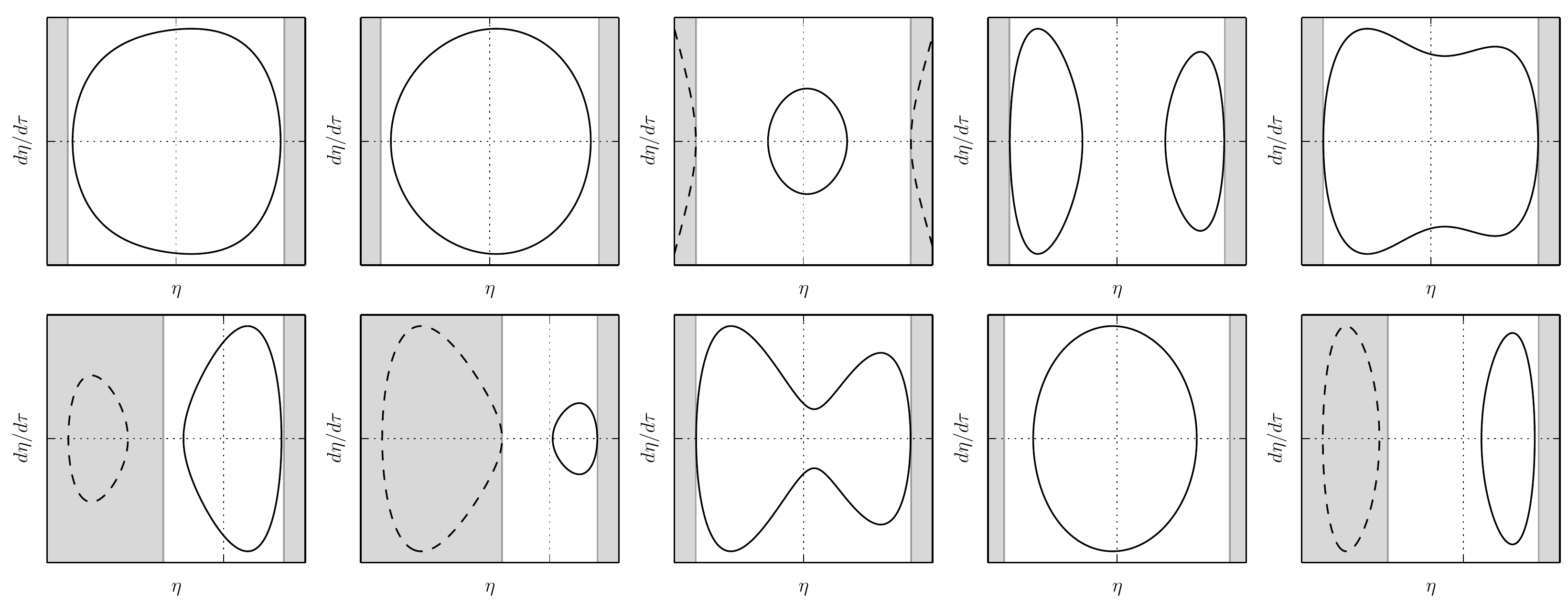}

\caption{Phase portraits for the $\eta$ variable in ten randomly-generated cases. The shaded regions corresponding to $\eta < -1$ and $\eta > 1$ are the domains
in which the $\eta$ variable does not represent any real physical coordinate. The trajectories in the $\left( \eta, d\eta/d\tau \right)$ plane
are elliptic curves.\label{fig:eta_poly}}
\end{figure*}

The equation of motion for $\eta$, eq. \eqref{eq:eta_tau}, reads:
\begin{equation}
\frac{d\eta}{d\tau} = -\frac{\eta^{2} p_{\eta}}{a^{2}} + \frac{p_{\eta}}{a^{2}}.\label{eq:eta_mot}
\end{equation}
Following the same procedure adopted for $\xi$, we can express $p_\eta$ by inverting eq. \eqref{eq:h_eta}, and rewrite
the equation of motion as
\begin{equation}
\frac{d\eta}{d\tau} = \pm \frac{\sqrt{f_\eta\left( \eta \right)}}{a^2},\label{eq:deta_dtau_00}
\end{equation}
where $f_\eta\left( \eta \right)$ is a polynomial of degree 4 in $\eta$:
\begin{multline}
f_\eta\left( \eta \right) = 2 \eta^{4} a^{2} h + \eta^{3} \left(- 2 \mu_{1} a + 2 \mu_{2} a\right) \\
+\eta^{2} \left(- 2 a^{2} h - 2 a^{2} h_{\eta}\right) + \eta \left(2 \mu_{1} a - 2 \mu_{2} a\right) \\
+ 2 a^{2} h_{\eta} - p_{\phi}^{2}.\label{eq:eta_poly}
\end{multline}
It is easy to verify by direct substitution that
\begin{equation}
f_\eta\left(\pm 1\right) = -p_\phi^2 < 0,
\end{equation}
which implies that in the domain of interest, $\eta\in\left(-1,1\right)$, the polynomial
$f_\eta\left( \eta \right)$ must have real roots.
Figure \ref{fig:eta_poly} visualises the phase portraits for $\eta$ in a few randomly-selected cases.

We can then adopt also for $\eta$ the simplified
inversion method from \citet[\S 20.6]{whittaker_course_1927}, and write the solution of eq. \eqref{eq:deta_dtau_00} as:
\begin{equation}
\eta\left( \tau \right) = \eta_r + \frac{1}{4}\frac{f_\eta^\prime\left( \eta_r \right)}{\wp_\eta\left( \frac{\tau - \tau_\eta}{a^2} \right)
- \frac{1}{24}f_\eta^{\prime\prime}\left( \eta_r \right)}.\label{eq:eta_sol}
\end{equation}
Analogously to $\xi_r$, $\eta_r$ is a root of the polynomial $f_\eta\left( \eta \right)$ and $\tau_\eta$ is the fictitious time
of root passage for the variable $\eta$.
The Weierstrassian function $\wp_\eta$ is defined in terms of the two invariants
\begin{align}
g_2 & = \mu_{1}^{2} a^{2} - 2 \mu_{1} \mu_{2} a^{2} + \mu_{2}^{2} a^{2} + \frac{a^{4} h^{2}}{3} + \frac{14 h}{3} a^{4} h_{\eta} \notag\\
&\quad + \frac{a^{4} h_{\eta}^{2}}{3} - 2 a^{2} h p_{\phi}^{2},\label{eq:g2_eta}\\
g_3 & = - \frac{\mu_{1}^{2} h}{3} a^{4} - \frac{\mu_{1}^{2} h_{\eta}}{3} a^{4} + \frac{\mu_{1}^{2} p_{\phi}^{2}}{4} a^{2} \notag\\ 
&\quad + \frac{2 \mu_{1}}{3} \mu_{2} a^{4} h + \frac{2 \mu_{1}}{3} \mu_{2} a^{4} h_{\eta} - \frac{\mu_{1} \mu_{2}}{2} a^{2} p_{\phi}^{2} \notag\\
&\quad - \frac{\mu_{2}^{2} h}{3} a^{4} - \frac{\mu_{2}^{2} h_{\eta}}{3} a^{4} + \frac{\mu_{2}^{2} p_{\phi}^{2}}{4} a^{2} \notag\\
&\quad + \frac{a^{6} h^{3}}{27} - \frac{11 h_{\eta}}{9} a^{6} h^{2} - \frac{11 h}{9} a^{6} h_{\eta}^{2} + \frac{a^{6} h_{\eta}^{3}}{27} \notag\\
&\quad + \frac{2 h^{2}}{3} a^{4} p_{\phi}^{2} + \frac{2 h}{3} a^{4} h_{\eta} p_{\phi}^{2},\label{eq:g3_eta}
\end{align}
and $\tau_\eta$ can be calculated as
\begin{equation}
\tau_\eta = a^2 \wp_\eta^{-1}\left[\frac{1}{24} f_\eta^{\prime\prime}\left( \eta_r \right) + \frac{f_\eta^\prime\left( \eta_r \right)}{4\left(\eta_0 - \eta_r\right)}\right].\label{eq:tau_eta_0}
\end{equation}
In analogy with the notation adopted for $\xi$, we refer to the real period of $\wp_\eta$ as $2\omega_\eta$. The period
of $\eta\left(\tau\right)$ is $2a^2\omega_\eta$.

\subsection{Solution for $\phi$ and for the time equation}
The equation of motion for $\phi$, eq. \eqref{eq:phi_tau}, and the time equation, eq. \eqref{eq:fic_time}, read:
\begin{align}
\frac{d\phi}{d\tau} & = \frac{p_\phi}{a^2} \left[ \frac{1}{\xi^2\left( \tau \right) - 1} + \frac{1}{1-\eta^2\left( \tau \right)}\right],\label{eq:expl_phi}\\
\frac{dt}{d\tau} & = \xi^2\left( \tau \right) - \eta^2\left( \tau \right).\label{eq:expl_t}
\end{align}
The solution of these two equations involves the integration by quadrature of a rational function of Weierstrass elliptic functions. In order reduce visual clutter,
we introduce the notation
\begin{align}
A_\xi & = f_\xi^\prime\left(\xi_r\right), & A_\eta & = f_\eta^\prime\left(\eta_r\right),\\
B_\xi & = \frac{1}{24}f_\xi^{\prime\prime}\left(\xi_r\right), & B_\eta & = \frac{1}{24}f_\eta^{\prime\prime}\left(\eta_r\right).\label{eq:Bxi}
\end{align}
$A_\xi$, $B_\xi$, $A_\eta$ and $B_\eta$ are all constants of motion depending on the physical parameters of the system and on the initial conditions.

The general theory of the integration of rational functions of Weierstrass elliptic functions is given, e.g., in \citet[Chapter VII]{halphen_traite_1886} 
and \citet[Chapter VII]{greenhill_applications_1959}. Here it will be enough
to first perform a fraction decomposition with respect to $\wp$, and then to integrate the decomposed fractions using the formul\ae{} from \citet[Chapter CXII]{jules_tannery_elements_1893}.
After the substitution of eqs. \eqref{eq:xi_sol} and \eqref{eq:eta_sol} into eq. \eqref{eq:expl_phi}, the decomposition for
$d\phi/d\tau$ is:
\begin{multline}
\frac{d\phi}{d\tau}  = \\
\frac{p_\phi A_{\xi}}{8 a^2 \left(\xi_{r} + 1\right)^2 \left[\wp_\xi\left( \frac{\tau - \tau_\xi}{a^2} \right) + \frac{A_{\xi} - 4 B_{\xi} \xi_{r} - 4 B_{\xi}}{4\left(\xi_r+1\right)}\right]} \\
- \frac{p_\phi A_{\xi}}{8 a^2 \left(\xi_{r} - 1\right)^2 \left[\wp_\xi\left( \frac{\tau - \tau_\xi}{a^2} \right) + \frac{A_{\xi} - 4 B_{\xi} \xi_{r} + 4 B_{\xi}}{4\left(\xi_r-1\right)}\right]} \\
- \frac{p_\phi A_{\eta}}{8 a^2 \left(\eta_{r} + 1\right)^2 \left[\wp_\eta\left( \frac{\tau - \tau_\eta}{a^2} \right) + \frac{A_{\eta} - 4 B_{\eta} \eta_{r} - 4 B_{\eta}}{4\left( \eta_r+1 \right)}\right]} \\
+ \frac{p_\phi A_{\eta}}{8 a^2 \left(\eta_{r} - 1\right)^2 \left[\wp_\eta\left( \frac{\tau - \tau_\eta}{a^2} \right) + \frac{A_{\eta} - 4 B_{\eta} \eta_{r} + 4 B_{\eta}}{4\left(\eta_r-1\right)}\right]} \\
+ \frac{p_\phi}{a^2\left(\xi_{r}^{2} - 1\right)} - \frac{p_\phi}{a^2\left(\eta_{r}^{2} - 1\right)}.\label{eq:phi_decomp}
\end{multline}
The decomposition for $dt/d\tau$ reads:
\begin{multline}
\frac{dt}{d\tau} = \frac{A_{\xi}^{2}}{16 \left[\wp_\xi\left( \frac{\tau - \tau_\xi}{a^2} \right) - B_{\xi}\right]^{2}} + \frac{A_{\xi} \xi_{r}}{2\left[\wp_\xi\left( \frac{\tau - \tau_\xi}{a^2} \right) - B_{\xi}\right]} + \xi_{r}^{2}\\
- \frac{A_{\eta}^{2}}{16 \left[\wp_\eta\left( \frac{\tau - \tau_\eta}{a^2} \right) - B_{\eta}\right]^{2}} - \frac{A_{\eta} \eta_{r}}{2\left[\wp_\eta\left( \frac{\tau - \tau_\eta}{a^2} \right)-B_{\eta}\right]} - \eta_{r}^{2}.\label{eq:t_decomp}
\end{multline}
In order to integrate eqs. \eqref{eq:phi_decomp} and \eqref{eq:t_decomp}, we employ two formul\ae{} from \citet[Chapter CXII]{jules_tannery_elements_1893} (see also \citet[\S 5.141]{gradshtein_table_2007}):
\begin{align}
\int\frac{du}{\wp\left(u\right) - \wp\left(v\right)} & = \frac{1}{\wp^\prime\left(v\right)}\left[\ln\frac{\sigma\left(u - v\right)}{\sigma\left(u + v\right)}+2u\zeta\left(v\right)\right],\label{eq:molk_00}\\
\int\frac{du}{\left[\wp\left(u\right) - \wp\left(v\right)\right]^2} & = \frac{1}{\wp^{\prime 2}\left(v\right)} \left[
-\zeta\left(u - v\right)-\zeta\left(u + v\right)
\vphantom{\int\frac{du}{\wp\left(u\right) - \wp\left(v\right)}}
\right.\notag\\
&\quad \left. - 2u\wp\left(v\right)-\wp^{\prime\prime}\left(v\right) \int\frac{du}{\wp\left(u\right) - \wp\left(v\right)}
\right].\label{eq:molk_01}
\end{align}
It can easily be recognized how the terms in eqs. \eqref{eq:phi_decomp} and \eqref{eq:t_decomp} are in the form of the integrands on the left-hand sides of eqs.
\eqref{eq:molk_00} and \eqref{eq:molk_01} (modulo additive and multiplicative constants)\footnote{It must be noted how the right-hand sides of eqs. \eqref{eq:molk_00} and \eqref{eq:molk_01} have singularities
when $\wp^\prime\left(v\right)$ is zero. In this situation, alternative formul\ae{} need to be used, as detailed in \citet[Chapter CXII]{jules_tannery_elements_1893} and \citet[\S 5.141]{gradshtein_table_2007}.
We will not consider this special case here.}\footnote{We need to point out a potential problem in the numerical evaluation of the right-hand side of eq. \eqref{eq:molk_00}, related to the appearance of the
multivalued complex logarithm. If one uses the principal value for $\ln$, the right-hand side of eq. \eqref{eq:molk_00} will be discontinuous in correspondence of the branch cuts. In Appendix A2 of
\citet{biscani_stark_2014} we discuss in detail the problem, and we provide a solution based on the Fourier series expansion of $\ln\sigma\left(z\right)$.}.

Following \citet[Chapter CXII]{jules_tannery_elements_1893}, we introduce the shorthand notation for the two integrals \eqref{eq:molk_00} and \eqref{eq:molk_01}
\begin{align}
\mathcal{J}_1\left(u,v\right) & = \int\frac{du}{\wp\left(u\right) - \wp\left(v\right)},\\
\mathcal{J}_2\left(u,v\right) & = \int\frac{du}{\left[\wp\left(u\right) - \wp\left(v\right)\right]^2}
\end{align}
(with the understanding that we will add a $\xi$ or $\eta$ subscript depending on the subscript of the Weierstrassian functions appearing in the integrals).
With this convention, we can integrate eq. \eqref{eq:phi_decomp} and obtain:
\begin{multline}
\phi\left(\tau\right) = \phi_0 \\
+\frac{p_\phi A_{\xi}}{8 \left(\xi_{r} + 1\right)^2}\left[\mathcal{J}_{\xi,1}\left(\frac{\tau-\tau_\xi}{a^2},v_{\xi,1}\right) - \mathcal{J}_{\xi,1}\left(-\frac{\tau_\xi}{a^2},v_{\xi,1}\right)\right] \\
-\frac{p_\phi A_{\xi}}{8 \left(\xi_{r} - 1\right)^2}\left[\mathcal{J}_{\xi,1}\left(\frac{\tau-\tau_\xi}{a^2},v_{\xi,2}\right) - \mathcal{J}_{\xi,1}\left(-\frac{\tau_\xi}{a^2},v_{\xi,2}\right)\right] \\
-\frac{p_\phi A_{\eta}}{8 \left(\eta_{r} + 1\right)^2}\left[\mathcal{J}_{\eta,1}\left(\frac{\tau-\tau_\eta}{a^2},v_{\eta,1}\right) - \mathcal{J}_{\eta,1}\left(-\frac{\tau_\eta}{a^2},v_{\eta,1}\right)\right] \\
+\frac{p_\phi A_{\eta}}{8 \left(\eta_{r} - 1\right)^2}\left[\mathcal{J}_{\eta,1}\left(\frac{\tau-\tau_\eta}{a^2},v_{\eta,2}\right) - \mathcal{J}_{\eta,1}\left(-\frac{\tau_\eta}{a^2},v_{\eta,2}\right)\right] \\
+\frac{p_\phi\tau}{a^2\left(\xi_{r}^{2} - 1\right)} - \frac{p_\phi\tau}{a^2\left(\eta_{r}^{2} - 1\right)},\label{eq:phi_tau_sol}
\end{multline}
where we have introduced the constants
\begin{align}
v_{\xi,1} & = \wp^{-1}\left[-\frac{A_{\xi} - 4 B_{\xi} \xi_{r} - 4 B_{\xi}}{4\left(\xi_r+1\right)}\right],\\
v_{\xi,2} & = \wp^{-1}\left[-\frac{A_{\xi} - 4 B_{\xi} \xi_{r} + 4 B_{\xi}}{4\left(\xi_r-1\right)}\right],\\
v_{\eta,1} & = \wp^{-1}\left[-\frac{A_{\eta} - 4 B_{\eta} \eta_{r} - 4 B_{\eta}}{4\left(\eta_r+1\right)}\right],\\
v_{\eta,2} & = \wp^{-1}\left[-\frac{A_{\eta} - 4 B_{\eta} \eta_{r} + 4 B_{\eta}}{4\left(\eta_r-1\right)}\right]
\end{align}
in order to simplify the notation. Analogously, the integration of eq. \eqref{eq:t_decomp} yields
\begin{multline}
t\left(\tau\right) = \\
\frac{a^2A_{\xi}^{2}}{16} \left[\mathcal{J}_{\xi,2}\left(\frac{\tau-\tau_\xi}{a^2},b_\xi\right) -  \mathcal{J}_{\xi,2}\left(-\frac{\tau_\xi}{a^2},b_\xi\right)\right]\\
+\frac{a^2A_{\xi}\xi_r}{2}\left[\mathcal{J}_{\xi,1}\left(\frac{\tau-\tau_\xi}{a^2},b_\xi\right) -  \mathcal{J}_{\xi,1}\left(-\frac{\tau_\xi}{a^2},b_\xi\right)\right]\\
-\frac{a^2A_{\eta}^{2}}{16} \left[\mathcal{J}_{\eta,2}\left(\frac{\tau-\tau_\eta}{a^2},b_\eta\right) -  \mathcal{J}_{\eta,2}\left(-\frac{\tau_\eta}{a^2},b_\eta\right)\right]\\
-\frac{a^2A_{\eta}\eta_r}{2}\left[\mathcal{J}_{\eta,1}\left(\frac{\tau-\tau_\eta}{a^2},b_\eta\right) -  \mathcal{J}_{\eta,1}\left(-\frac{\tau_\eta}{a^2},b_\eta\right)\right]\\
+\left(\xi_r^2-\eta_r^2\right)\tau,\label{eq:t_sol}
\end{multline}
where we have introduced the constants
\begin{align}
b_\xi & = \wp_\xi^{-1}\left(B_\xi\right),\label{eq:bxi}\\
b_\eta & = \wp_\eta^{-1}\left(B_\eta\right)
\end{align}
again in order to simplify the notation.

\subsection{Solution for $p_\xi$ and $p_\eta$}
The explicit solution in fictitious time for $p_\xi$ and $p_\eta$ can be computed by inverting eqs. \eqref{eq:xi_mot} and \eqref{eq:eta_mot} to yield
\begin{align}
p_\xi\left(\tau\right) & = \frac{a^2}{\xi^2\left(\tau\right)-1}\frac{d\xi\left(\tau\right)}{d\tau},\label{eq:pxi_sol}\\
p_\eta\left(\tau\right) & = \frac{a^2}{1-\eta^2\left(\tau\right)}\frac{d\eta\left(\tau\right)}{d\tau}.\label{eq:peta_sol}
\end{align}
The derivatives of $\xi\left(\tau\right)$ and $\eta\left(\tau\right)$ can be computed from the solutions \eqref{eq:xi_sol} and \eqref{eq:eta_sol}:
\begin{align}
\frac{d\xi\left(\tau\right)}{d\tau} & = -\frac{A_\xi\wp_\xi^\prime\left( \frac{\tau - \tau_\xi}{a^2} \right)}
{4a^2\left[\wp_\xi\left( \frac{\tau - \tau_\xi}{a^2} \right) - B_\xi\right]^2},\\
\frac{d\eta\left(\tau\right)}{d\tau} & = -\frac{A_\eta\wp_\eta^\prime\left( \frac{\tau - \tau_\eta}{a^2} \right)}
{4a^2\left[\wp_\eta\left( \frac{\tau - \tau_\eta}{a^2} \right) - B_\eta\right]^2}.
\end{align}

\section{Analysis of the results}
\label{sec:analysis}
In the previous sections we computed the full exact solution of the three-dimensional E3BP. The evolution
of the coordinates $\xi$, $\eta$ and $\phi$ in fictitious time is given by eqs. \eqref{eq:xi_sol}, \eqref{eq:eta_sol}
and \eqref{eq:phi_tau_sol}. The evolution in fictitious time of the momenta $p_\xi$ and $p_\eta$ is given by eqs.
\eqref{eq:pxi_sol} and \eqref{eq:peta_sol} (the third momentum, $p_\phi$, is a constant of motion). The connection
between the fictitious time $\tau$ and the real time $t$ is given by eq. \eqref{eq:t_sol}. The equations are valid
for all initial conditions and physical parameters of the three-dimensional system (that is, when $p_\phi$ is not zero).

The solutions for $\xi$, $\eta$, $p_\xi$ and $p_\eta$ are expressed solely in terms of the elliptic functions $\wp$ and
$\wp^\prime$. $\xi$ and $p_\xi$ are thus periodic functions of $\tau$ with period $2a^2\omega_\xi$, while $\eta$ and $p_\eta$
are periodic functions of $\tau$ with period $2a^2\omega_\eta$. In general, the two periods will be different.

The solutions for $\phi\left( \tau \right)$ and $t\left( \tau \right)$ involve also the Weierstrassian functions $\sigma$
and $\zeta$ (see eqs. \eqref{eq:molk_00} and \eqref{eq:molk_01}), which are \emph{not} elliptic functions: they are \emph{quasi-periodic}
functions\footnote{Here we use the term ``quasi-periodic'' in the following sense:
a function $f$ is quasi-periodic with quasi-period $T$ if $f\left(z + T \right) = g\left(z,f\left(z\right)\right)$.
In the specific cases of $\sigma$ and $\zeta$, the following relations hold:
\begin{align}
\sigma\left( z + T \right) & = A\mathrm{e}^{zB}\sigma\left( z \right),\\
\zeta\left(z + T \right) & = \zeta\left(z \right) + C,
\end{align}
where $A$, $B$ and $C$ are constants \citep[eqs. 18.2.19 and 18.2.20]{abramowitz_handbook_1964}.
}.
The fictitious time derivatives \eqref{eq:expl_phi} and \eqref{eq:expl_t} of $\phi$ and $t$, however, involve only the
function $\wp$ and they can be thus seen as sums of functions with two different periods, $2a^2\omega_\xi$ and $2a^2\omega_\eta$.
Such functions are sometimes called \emph{almost-periodic} functions \citep{besicovitch_almost_1932}.

In a way, the fictitious time $\tau$ can be considered as the E3BP analogue of the eccentric anomaly in the two-body problem.
Kepler's equation for the
elliptic two-body problem reads:
\begin{equation}
t\left( E \right) = \frac{E-e\sin E}{n},
\end{equation}
where $E$ is the eccentric anomaly, $e$ the eccentricity of the orbit and $n$ the (constant) mean motion. Clearly, Kepler's
equation is structurally similar to (albeit much simpler than) eq. \eqref{eq:t_sol}: they are both transcendental equations featuring a combination of linear
and periodic parts. Kepler's equation is a quasi-periodic function, so that
\begin{equation}
t\left( E + 2\pi \right) = t\left( E \right) + \frac{2\pi}{n}.
\end{equation}
In a similar fashion, $t \left( \tau \right)$ is the sum of two parts, one quasi-periodic with quasi-period $2a^2\omega_\xi$, the other
quasi-periodic with quasi-period $2a^2\omega_\eta$. Following the nomenclature introduced earlier, we can then refer to 
$t\left( \tau \right)$ (and to $\phi\left( \tau \right)$ as well, since it is structurally identical)
as an \emph{almost quasi-periodic} function. In \S \ref{subsec:period} we will examine periodicity
and quasi-periodicity in the E3BP in more detail.

\subsection{Boundedness and regions of motion}
As we noted in the previous paragraph, the solution of the E3BP in terms of Weierstrassian functions is expressed as a set of
unique formul\ae{} valid for all initial conditions and physical parameters of the system. That is, both bounded and unbounded
orbits can be described by the same equations. Now, according to eqs. \eqref{eq:rho_ell} and \eqref{eq:z_ell}, the test particle can go to infinity only when $\xi$ goes to infinity
(by definition, the $\eta$ coordinate is confined to the $\left[-1,1\right]$ interval). The solution for
$\xi\left(\tau\right)$, here reproduced for convenience,
\begin{equation}
\xi\left( \tau \right) = \xi_r + \frac{1}{4}\frac{f_\xi^\prime\left( \xi_r \right)}{\wp_\xi\left( \frac{\tau - \tau_\xi}{a^2} \right)
- \frac{1}{24}f_\xi^{\prime\prime}\left( \xi_r \right)},\label{eq:xi_sol2}
\end{equation}
shows how the necessary and sufficient condition for the motion to be bounded is
\begin{equation}
\wp_{\xi,\textnormal{min}} > \frac{1}{24}f_\xi^{\prime\prime}\left( \xi_r \right),\label{eq:bounded_cond}
\end{equation}
where $\wp_{\xi,\textnormal{min}}$ is the minimum value assumed by $\wp_\xi$ in
the real period $2\omega_\xi$. When this condition holds, the right-hand side of eq. \eqref{eq:xi_sol2} has no
poles and $\xi\left( \tau \right)$ is bounded. Vice versa, when the condition does not hold $\xi$ will reach infinity
in a finite amount of fictitious time. According to the theory of elliptic functions, within the real period and
on the real axis
\begin{equation}
\wp_{\xi,\textnormal{min}} = \wp_\xi\left( \omega_\xi \right).\label{eq:P_min}
\end{equation}
That is, the global minimum of $\wp$ on the real axis is in correspondence of the real half-period. In addition,
\begin{equation}
\wp_\xi\left( \omega_\xi \right) = e_i,\label{eq:om_e}
\end{equation}
where $e_i$ is one of the roots of the Weierstrass cubic \eqref{eq:cubic_wp}. By using eqs. \eqref{eq:om_e}
and \eqref{eq:P_min}, we can thus rewrite the condition \eqref{eq:bounded_cond} as
\begin{equation}
e_i > \frac{1}{24}f_\xi^{\prime\prime}\left( \xi_r \right).
\end{equation}

For unbounded orbits, we can compute the fictitious time at which $\xi$ goes to infinity, $\tau_\infty$, using the condition:
\begin{equation}
\wp_\xi\left( \frac{\tau_\infty - \tau_\xi}{a^2} \right)
- \frac{1}{24}f_\xi^{\prime\prime}\left( \xi_r \right) = 0,
\end{equation}
that is,
\begin{equation}
\tau_\infty = \tau_\xi + a^2\wp^{-1}\left[ \frac{1}{24}f_\xi^{\prime\prime}\left(\xi_r\right) \right].\label{eq:tau_infty}
\end{equation}
When the motion is unbounded $\tau_\infty$ is a real quantity, whereas when the motion is bounded $\tau_\infty$ becomes complex.
With the definitions \eqref{eq:bxi} and \eqref{eq:Bxi}, we can rewrite eq. \eqref{eq:tau_infty} as
\begin{equation}
\tau_\infty = \tau_\xi + a^2\wp_\xi^{-1}\left( B_\xi\right).
\end{equation}
We can then immediately verify how the substitution of $\tau_\infty$ for $\tau$ in eq. \eqref{eq:t_decomp} leads to the two denominators
of the form
\begin{equation}
\wp_\xi\left( \frac{\tau - \tau_\xi}{a^2} \right) - B_{\xi}
\end{equation}
on the right-hand side to go to zero. That is, for $\tau = \tau_\infty$ $dt/d\tau$ has a vertical asymptote, and the real time thus goes to
infinity in a finite amount of fictitious time.

In contrast to $t\left(\tau\right)$, $\phi\left(\tau\right)$ has no asymptotes in fictitious time. For $\tau = \tau_\infty$,
in unbounded orbits $\phi\left(\tau\right)$ assumes the finite value $\phi_\infty$ (which can be calculated via the substitution of $\tau_\infty$ for $\tau$ in eq.
\eqref{eq:phi_tau_sol}). In the E3BP, $\phi_\infty$ plays the same role that the approach and departure angles play in hyperbolic trajectories
in the two-body problem.

\begin{figure*}
\includegraphics[width=1\textwidth]{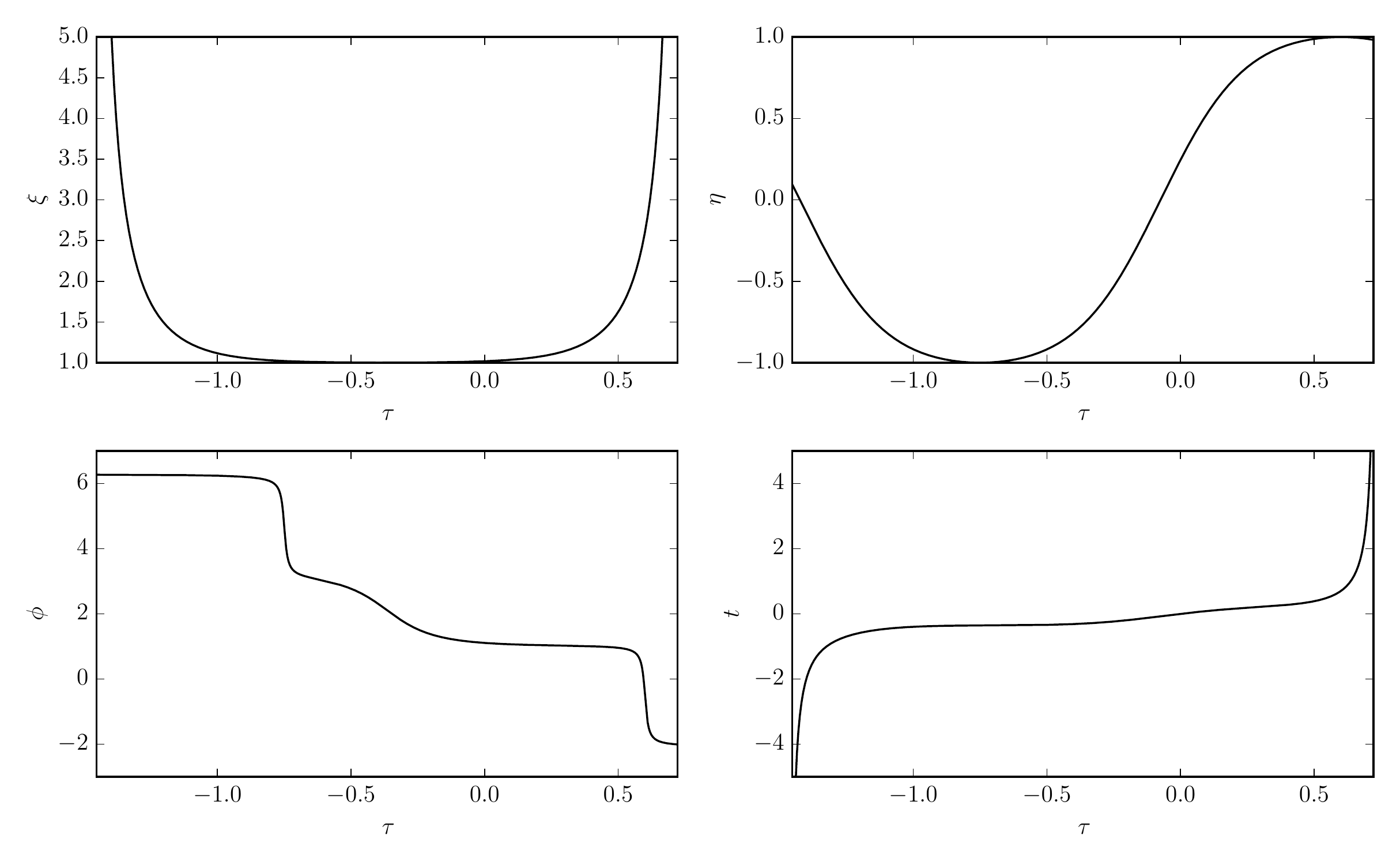}

\caption{Evolution in fictitious time $\tau$ of the coordinates $\xi$, $\eta$ and $\phi$, and of the real time $t$ in a representative
unbounded orbit in the E3BP. $\xi\left(\tau\right)$ features two vertical asymptotes, reaching infinity in a finite amount of fictitious time. $t\left(\tau\right)$
also has two vertical asymptotes. $\eta\left(\tau\right)$ is bounded in the interval $\left[-1,1\right]$ by its geometrical definition,
and $\phi\left(\tau\right)$ assumes finite values when $\xi$ and $t$ go to infinity.\label{fig:unbounded}}
\end{figure*}

It should be noted that $\wp^{-1}$ is a multivalued function, and there are thus two possible values to choose from
for $\tau_\infty$ within the fundamental parallelogram of $\wp$. This duality physically corresponds to the fact that unbounded orbits have two
asymptotes, one inbound and one outbound. Correspondingly, there are two possible $\phi_\infty$ values, one at
$t=-\infty$ and the other one at $t=\infty$. Figure \ref{fig:unbounded} illustrates graphically the evolution in fictitious time of the coordinates and of $t$
in a representative unbounded trajectory.

It is worth stressing that, from a purely mathematical point of view, even in unbounded trajectories the evolution of $\xi$ is still periodic.
This periodicity is in fictitious time and it does not carry over to the evolution in real time because of the asymptotes in $t\left(\tau\right)$.

\begin{figure*}
\includegraphics[width=1\textwidth]{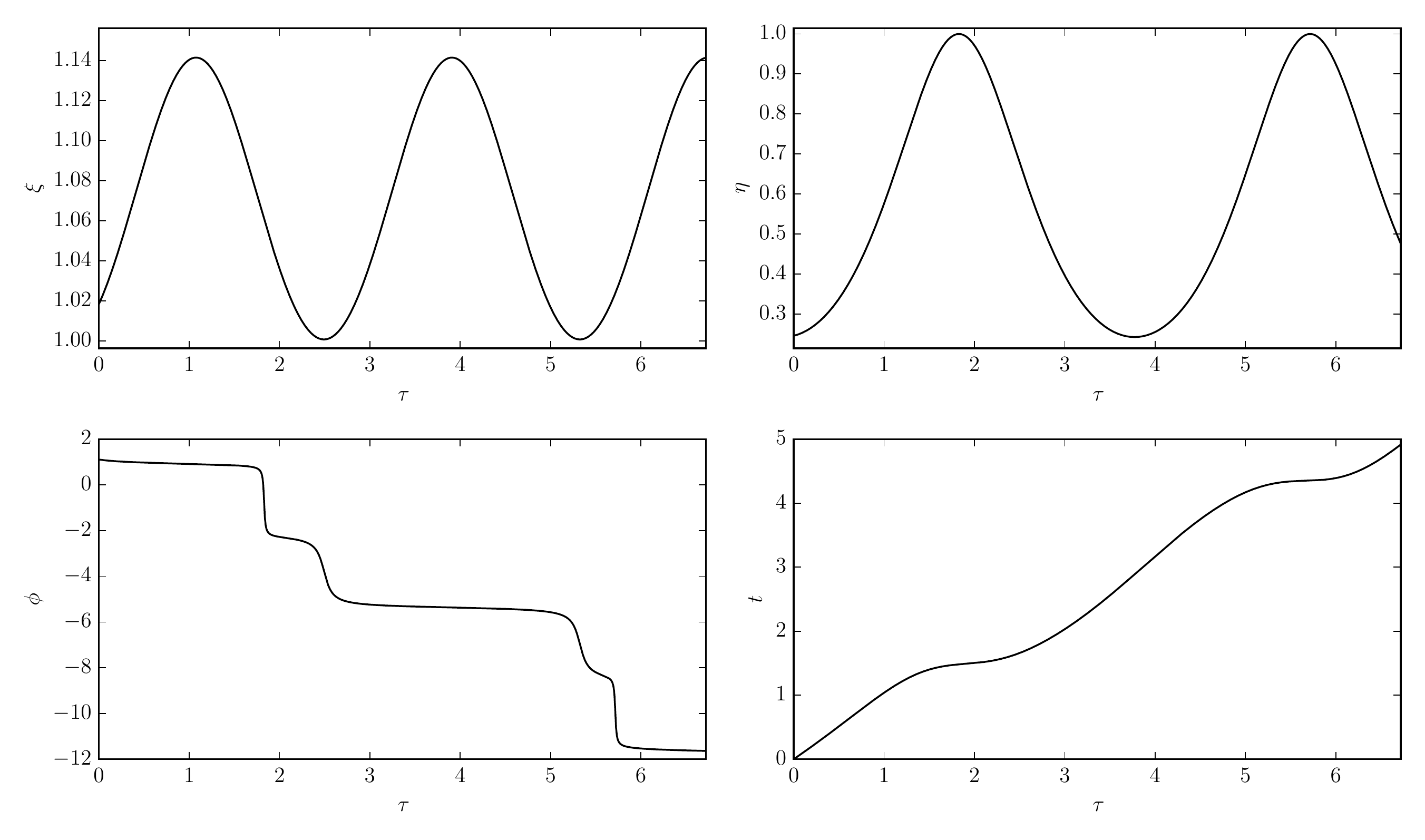}

\caption{Evolution in fictitious time $\tau$ of the coordinates $\xi$, $\eta$ and $\phi$, and of the real time $t$ in a representative
bounded orbit in the E3BP. $\xi\left(\tau\right)$ and $\eta\left(\tau\right)$ are periodic functions with two different periods. $t\left(\tau\right)$
and $\phi\left(\tau\right)$ are almost quasi-periodic functions.\label{fig:bounded}}
\end{figure*}

In bounded trajectories, by contrast, both $\xi$ and $\eta$ vary periodically within the finite ranges $\left[\xi_\textnormal{min},\xi_\textnormal{max}\right]$ and
$\left[\eta_\textnormal{min},\eta_\textnormal{max}\right]$. Consequently, $t\left(\tau\right)$ has no asymptotes in bounded trajectories, and the periodicity of
$\xi$ and $\eta$ in fictitious time translates to a periodicity in real time. Figure \ref{fig:bounded} illustrates graphically the evolution in fictitious time of the coordinates and of $t$
in a representative bounded trajectory.

\begin{figure}
\includegraphics[width=.47\textwidth]{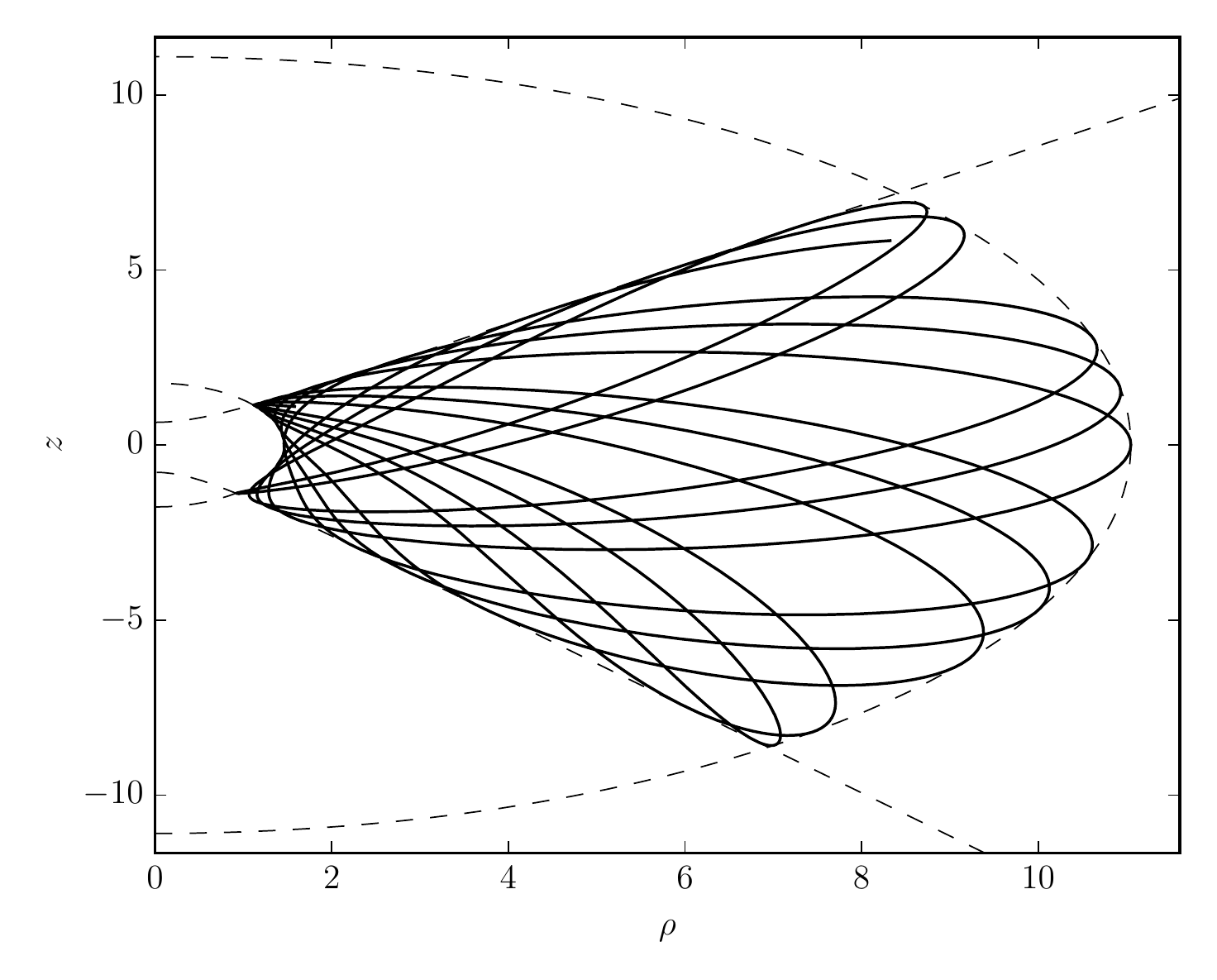}

\caption{Regions of motion on the $\left(\rho,z\right)$ plane in a representative bounded trajectory. The motion is confined within the two ellipses implicitly
defined by $\xi\left(\rho,z\right) = \left\{\xi_\textnormal{min},\xi_\textnormal{max}\right\}$ and within the two hyperbolic branches defined implicitly by
$\eta\left(\rho,z\right) = \left\{\eta_\textnormal{min},\eta_\textnormal{max}\right\}$. Both the ellipses and the hyperbolic branches are represented by the dashed lines.
The solid line line represents the actual trajectory of the test particle. The two centres of attraction are lying on the $z$ axis at $z=\pm1$.\label{fig:regions}}
\end{figure}

In Section \ref{sec:formulation}, we introduced the elliptic-cylindrical coordinate system defined by eqs. \eqref{eq:xidef} and \eqref{eq:etadef}:
\begin{align}
\xi & = \frac{\sqrt{\rho^2+\left( z + a\right)^2} + \sqrt{\rho^2+\left( z - a\right)^2}}{2a},\\
\eta & = \frac{\sqrt{\rho^2+\left( z + a\right)^2}-\sqrt{\rho^2+\left( z - a\right)^2}}{2a}.
\end{align}
For fixed values of $\xi$ and $\eta$, these two equations define confocal ellipses and hyperbolic branches in the $\left(\rho,z\right)$ plane. This means that,
in bounded orbits, the motion of the test particle in the $\left(\rho,z\right)$ plane is confined in the intersection of two geometric regions:
\begin{itemize}
\item an elliptic ring implicitly defined by the condition $\xi_\textnormal{min}\leq\xi\leq\xi_\textnormal{max}$,
\item a hyperbolic ring implicitly defined by the condition $\eta_\textnormal{min}\leq\eta\leq\eta_\textnormal{max}$.
\end{itemize}
The regions of motion in a representative bounded case are illustrated in Figure \ref{fig:regions}.

We can explicitly compute the fictitious time at which $\xi$ and $\eta$ assume their minimum and maximum values. According to the theory of elliptic functions,
$\wp_\xi$ has a global finite minimum on the real axis in correspondence of the real half-period $\omega_\xi$ (and, analogously, $\wp_\eta$ has a global minimum at
$\omega_\eta$). The global maximum for $\wp_\xi$ on the real axis is at $2k\omega_\xi$ ($k\in\mathbb{Z}$), where the value of the function is $+\infty$. Consequently:
\begin{align}
\xi_\textnormal{min} & = \xi\left( \tau_\xi \right), &  \xi_\textnormal{max} & = \xi\left( a^2\omega_\xi+\tau_\xi \right),\\
\eta_\textnormal{min} & = \eta\left( \tau_\eta \right), & \eta_\textnormal{max} & = \eta\left( a^2\omega_\eta+\tau_\eta \right).
\end{align}

\subsection{Periodicity and quasi-periodicity}
\label{subsec:period}
In the previous sections we have mentioned how the evolution of $\xi$ and $\eta$ in fictitious time is periodic, with two periods that,
in general, will be different. The periodicity in fictitious time of each coordinate translates to a periodicity in real time only for
bounded orbits. By contrast, the evolution in fictitious time of the third coordinate $\phi$ and of the real time $t$ derives from the integration
of almost periodic functions, and it is thus almost quasi-periodic. We are now going to examine in more detail the behaviour of $\phi$ and $t$. We will
focus on the study of $\phi\left(\tau\right)$, as $t\left(\tau\right)$ is structurally identical.

\begin{figure}
\includegraphics[width=.47\textwidth]{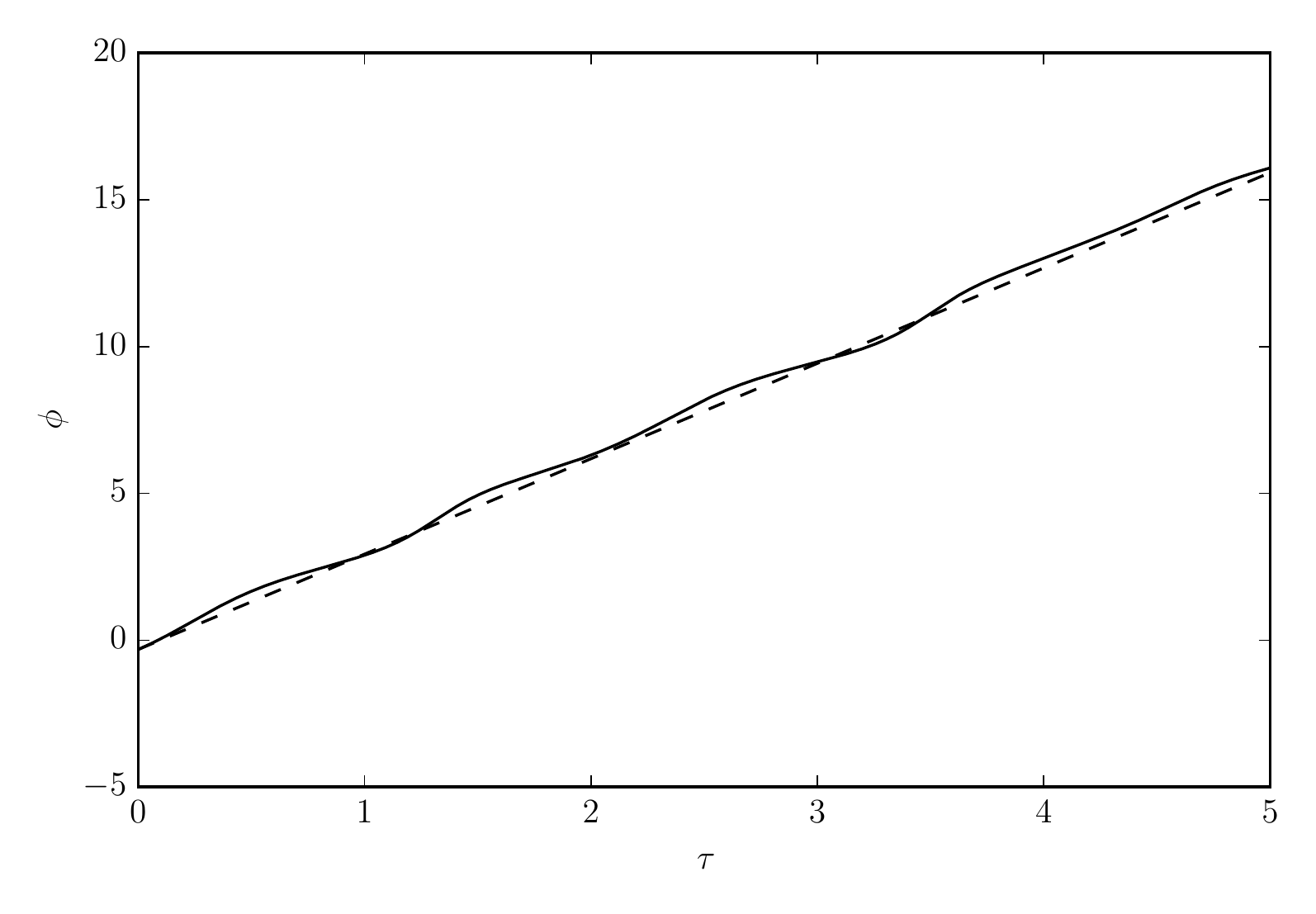}

\caption{Evolution in fictitious time $\tau$ of the $\phi$ coordinate in a representative bounded trajectory in the E3BP. The solid line
represents the actual evolution of the $\phi$ coordinate, while the dashed line represents the average rate of change of $\phi$
in fictitious time, which amounts to $\Phi_\xi+\Phi_\eta$ (see eq. \eqref{eq:av_phi_rate}). $\phi$ is an almost quasi-periodic function
of $\tau$.\label{fig:av_phi}}
\end{figure}

The derivative in fictitious time for $\phi\left(\tau\right)$, eq. \eqref{eq:expl_phi},
can be seen as a linear combination of two periodic functions:
\begin{equation}
\frac{d\phi}{d\tau} = \frac{d\phi_\xi\left(\tau\right)}{d\tau} + \frac{d\phi_\eta\left(\tau\right)}{d\tau},
\end{equation}
where
\begin{align}
\frac{d\phi_\xi\left(\tau\right)}{d\tau} & = \frac{p_\phi}{a^2\left[\xi^2\left( \tau \right) - 1\right]},\\
\frac{d\phi_\eta\left(\tau\right)}{d\tau} & = \frac{p_\phi}{a^2\left[1-\eta^2\left( \tau \right)\right]}.
\end{align}
The sign of $d\phi/d\tau$ is either always positive or always negative, depending on the sign of the constant $p_\phi$,
and thus $\phi\left(\tau\right)$ is a monotonic function.
$\phi_\xi\left(\tau\right)$ is given by the terms in eq. \eqref{eq:phi_tau_sol} related to $\xi$,
and similarly $\phi_\eta\left(\tau\right)$ is given by the terms related to $\eta$, so that eq. \eqref{eq:phi_tau_sol}
can be written as
\begin{equation}
\phi\left(\tau\right) = \phi_0 + \phi_\xi\left(\tau\right) + \phi_\eta\left(\tau\right).
\end{equation}
Since their derivatives are periodic functions with periods $2\omega_\xi$ and $2\omega_\eta$, $\phi_\xi\left(\tau\right)$
and $\phi_\eta\left(\tau\right)$ are arithmetic quasi-periodic functions with quasi-periods $2\omega_\xi$ and $2\omega_\eta$.
That is, for any $\tau\in\mathbb{R}$,
\begin{align}
\frac{\phi_\xi\left(\tau+2\omega_\xi\right)-\phi_\xi\left(\tau\right)}{2\omega_\xi} & = \Phi_\xi,\\
\frac{\phi_\eta\left(\tau+2\omega_\eta\right)-\phi_\eta\left(\tau\right)}{2\omega_\eta} & = \Phi_\eta,
\end{align}
where $\Phi_\xi$ and $\Phi_\eta$ are constants representing the average rate of change of $\phi_\xi\left(\tau\right)$
and $\phi_\eta\left(\tau\right)$. It follows then that the average rate of change of $\phi\left(\tau\right)$ is simply
\begin{equation}
\Phi_\xi+\Phi_\eta.
\end{equation}
That is, $\phi\left(\tau\right)$ will oscillate almost-periodically around a line parallel to the line
\begin{equation}
\left(\Phi_\xi+\Phi_\eta\right)\tau.\label{eq:av_phi_rate}
\end{equation}
The situation is illustrated in Figure \ref{fig:av_phi}.

The ratio between the periods of $\xi$ and $\eta$, $\omega_\xi/\omega_\eta$, is in general a real value. If
\begin{equation}
\frac{\omega_\xi}{\omega_\eta} = \frac{n}{m},
\end{equation}
where $n$ and $m$ are two coprime positive integers, then $\xi$ and $\eta$ will share the same finite period $T = 2m\omega_\xi= 2n\omega_\eta$:
\begin{align}
\xi\left(\tau\right) & = \xi\left(\tau + 2m\omega_\xi\right) = \xi\left(\tau + T\right),\\
\eta\left(\tau\right) & = \eta\left(\tau + 2n\omega_\eta\right) = \eta\left(\tau + T\right).
\end{align}
We refer to this case as an \emph{isochronous} configuration. In an isochronous configuration, $d\phi/d\tau$ and $dt/d\tau$ become periodic functions
of period $T$.
$\phi\left(\tau\right)$ and $t\left(\tau\right)$ are integrals of periodic functions, and thus they are arithmetic quasi-periodic functions:
\begin{align}
\phi\left(\tau + T\right) & = \phi\left(\tau\right) + \phi_T,\\
t\left(\tau + T\right) & = t\left(\tau\right) + t_T,
\end{align}
where $\phi_T$ and $t_T$ are two constants depending only on the initial conditions and physical parameters of the system.
From a geometric point of view, an isochronous configuration generates a quasi-periodic three-dimensional trajectory. That is,
after a quasi period $T$, the coordinates $\xi$ and $\eta$ and the momenta $p_\xi$ and $p_\eta$ will assume again their original
values, while the coordinate $\phi$ will be augmented by $\phi_T$.

\begin{figure*}
\includegraphics[width=\textwidth]{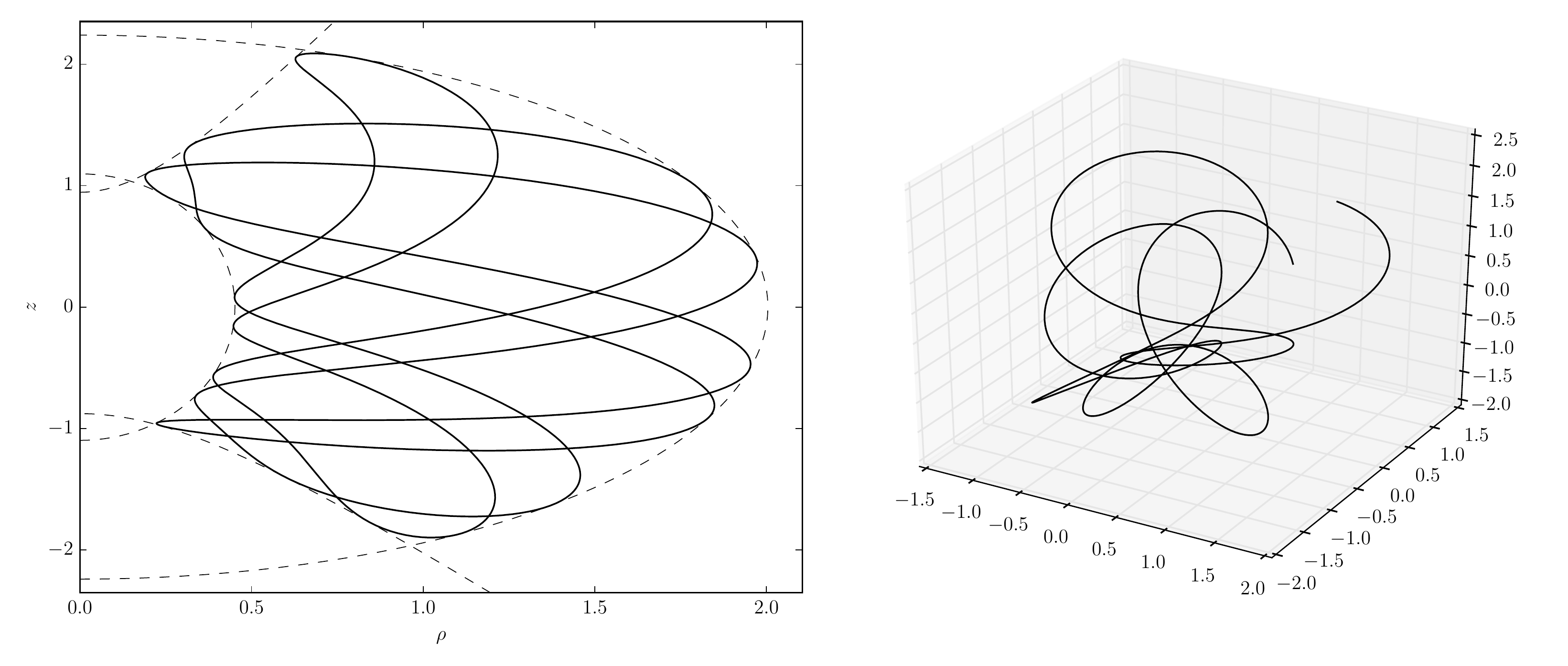}

\caption{Graphical depiction of a representative quasi-periodic orbit in the E3BP over a single quasi-period. In this specific case, the periods of $\xi$ and $\eta$ are in a ratio of $3/7$ with an accuracy
of $10^{-14}$. The left panel displays the trajectory in the $\left( \rho, z \right)$ plane (solid line), together with the boundaries of the region of motion (dashed lines).
The right panel displays the trajectory in the three-dimensional space. In quasi-periodic orbits, the trajectory in the $\left( \rho, z \right)$ plane is a closed
figure similar to a Lissajous curve (the difference being that here the periodic functions defining the figure are elliptic functions, rather than circular functions). In the
three-dimensional space the trajectory is not closed: after one quasi-period $\xi$ and $\eta$ will assume again the original values, while $\phi$ will be augmented
by a constant quantity.\label{fig:qperiodic}}
\end{figure*}

It is possible to look for isochronous configurations by setting up a numerical search. After fixing two coprime positive integers
$n$ and $m$ and the physical parameters of the system, the goal will be to find a set of initial conditions that minimises the quantity
\begin{equation}
\left| m\omega_\xi - n\omega_\eta \right|.
\end{equation}
Such a search can be performed with standard minimisation algorithms. In this case, we used the SLSQP algorithm from \citet{kraft_algorithm_1994}, as implemented
in the {\sc SciPy} Python library \citep{scipy}.
Figure \ref{fig:qperiodic} displays a representative quasi-periodic trajectory found by a numerical search.

\begin{figure*}
\includegraphics[width=\textwidth]{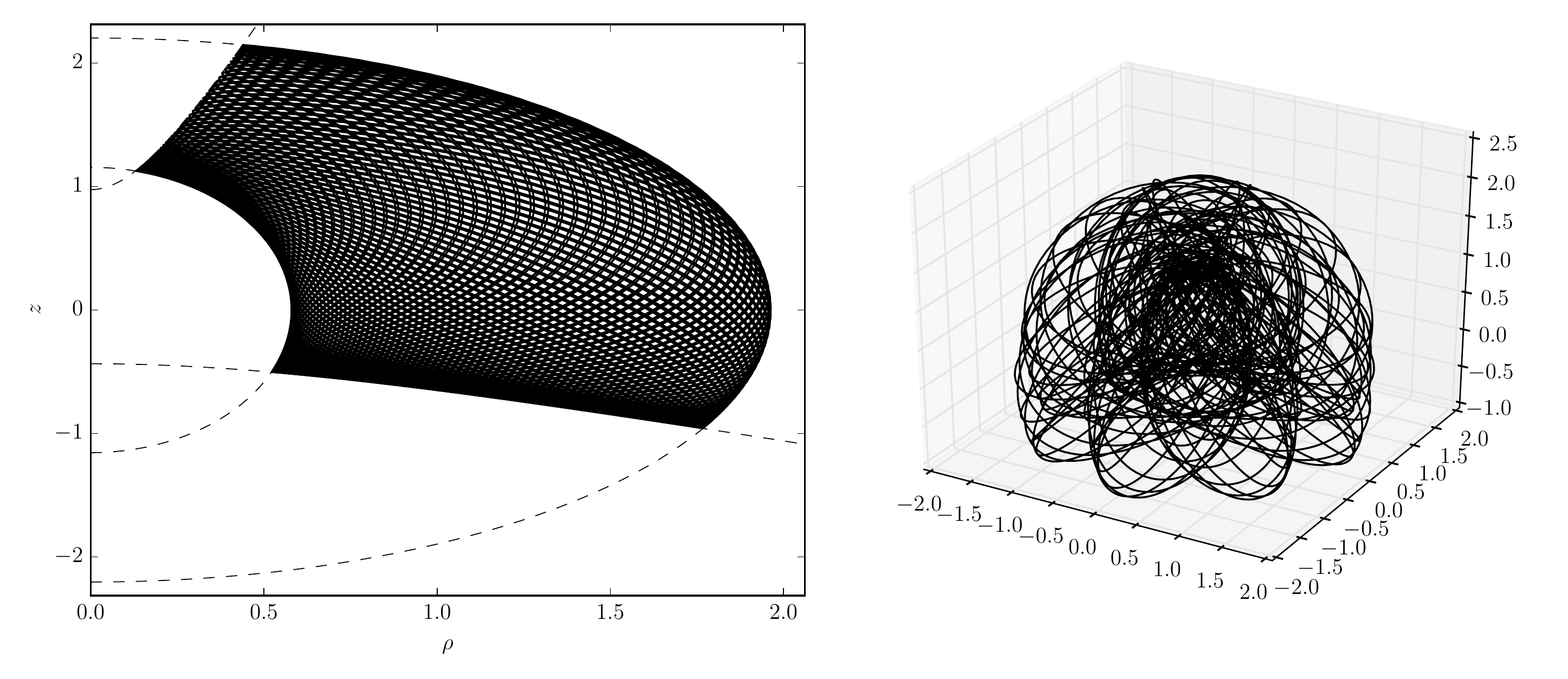}

\caption{Graphical depiction of a representative periodic orbit in the E3BP over a single period. In this specific case, the periods of $\xi$ and $\eta$ are in a ratio of $91/99$
with an accuracy of $10^{-13}$ and $\frac{\phi_T}{2\pi}=96$ with an accuracy of $10^{-11}$. The left panel displays the trajectory in the $\left( \rho, z \right)$ plane (solid line),
together with the boundaries of the region of motion (dashed lines).
The right panel displays the trajectory in the three-dimensional space. Analogously to a quasi-periodic orbit, the trajectory in the $\left( \rho, z \right)$ plane is a closed elliptic
Lissajous figure. Unlike in the quasi-periodic case, the three-dimensional trajectory is also a closed figure.
For this orbit, $a=1$, $\mu_1=1$, $\mu_2=.05$, the initial cartesian posision vector is $\left(1.20793759666736,-0.493320558636725,1.19760678594565\right)$,
and the initial cartesian velocity vector is $\left(-0.498435147674914,0.548228167205306,0.496626916283632\right)$. The period is $405.074289498234$.\label{fig:periodic}}
\end{figure*}

As a next step, we can look for periodic orbits: if, in an isochronous configuration, the $\phi_T$ constant is commensurable with $2\pi$, then the trajectory of the test particle
in the three-dimensional space will be a closed curve. In other words, the search for periodic orbits can be cast as the minimisation of the function
\begin{equation}
\left| m\omega_\xi - n\omega_\eta \right| + \left| \phi_T \% \frac{2\pi}{k} \right|,
\end{equation}
where $n$, $m$ and $k$ are positive integers, $n$ and $m$ are coprime and $\%$ is the modulo operator. Figure \ref{fig:periodic} displays a representative periodic trajectory found by a numerical search.

\subsection{Equilibrium points}
In the last section of our analysis, we will briefly examine the equilibrium points in the E3BP. It is clear from the form of the Lagrangian \eqref{eq:lagr00} that in cartesian coordinates
there is an equilibrium point lying on the $z$ axis, where the forces exerted by the two bodies are balanced. This is a ``real'' equilibrium point, in the sense
that the test particle will be at rest if placed in this point with a null initial velocity. This equilibrium point corresponds to the $\textnormal{L}_{1,2,3}$ Lagrangian points in the circular restricted
three-body problem.

In elliptic-cylindrical coordinates, we cannot properly characterise the equilibrium point on the $z$ axis as it lies in correspondence of a singularity of the coordinate system. On the other
hand, in elliptic-cylindrical coordinates we have another set of equilibrium points characterised by constant $\xi$ and/or $\eta$. These are not equilibrium points in which the particle is at rest:
\begin{itemize}
 \item if only $\xi$ is constant, then the motion is confined to a section of the surface of an ellipsoid of revolution,
 \item if only $\eta$ is constant, then the motion is confined to a section of the surface of a hyperboloid of revolution,
 \item if both $\xi$ and $\eta$ are constant, the motion is confined to a circle resulting from the intersection of an ellipsoid and a hyperboloid of revolution.
\end{itemize}
The third case, in particular, corresponds to the following initial setup:
\begin{itemize}
 \item the net force acting on the particle is parallel to the $xy$ plane,
 \item the velocity vector of the particle is also parallel to the $xy$ plane,
 \item the direction and the magnitude of the velocity vector are those of a Keplerian circular orbit with a virtual centre of attraction
 on the $z$ axis whose mass is exerting a force equal to the net force acting on the particle.
\end{itemize}
In other words, this case corresponds to a circular orbit parallel to the $xy$ plane along which the $z$ components of the forces from the two centres of attraction
cancel each other, leaving a net force directed towards the $z$ axis. We refer to this particular setup as a \emph{displaced circular orbit}\footnote{Displaced circular orbits
are also present in the Stark problem, which is a limiting case of the gravitational E3BP \citep{namouni_accelerated_2007,lantoine_complete_2011,biscani_stark_2014}}.

\begin{figure*}
\includegraphics[width=\textwidth]{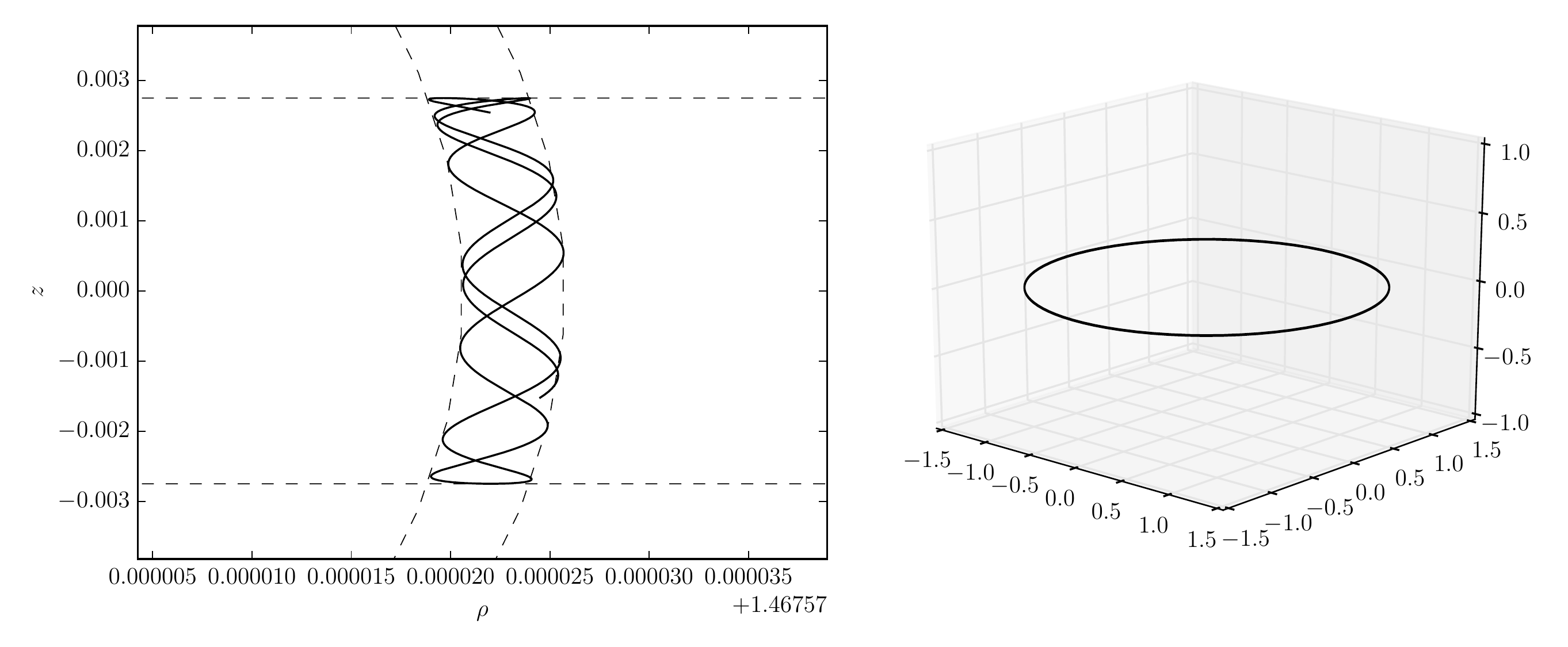}

\caption{Graphical depiction of a slightly perturbed displaced circular orbit in the E3BP near the $xy$ plane. The initial conditions have been chosen close to the setup described by
eqs. \eqref{eq:d_z_bal} and \eqref{eq:v_d}. The left panel displays the trajectory in the $\left( \rho, z \right)$ plane (solid line),
together with the boundaries of the region of motion (dashed lines). The right panel displays the trajectory in the three-dimensional space.
\label{fig:spring}}
\end{figure*}

\begin{figure*}
\includegraphics[width=\textwidth]{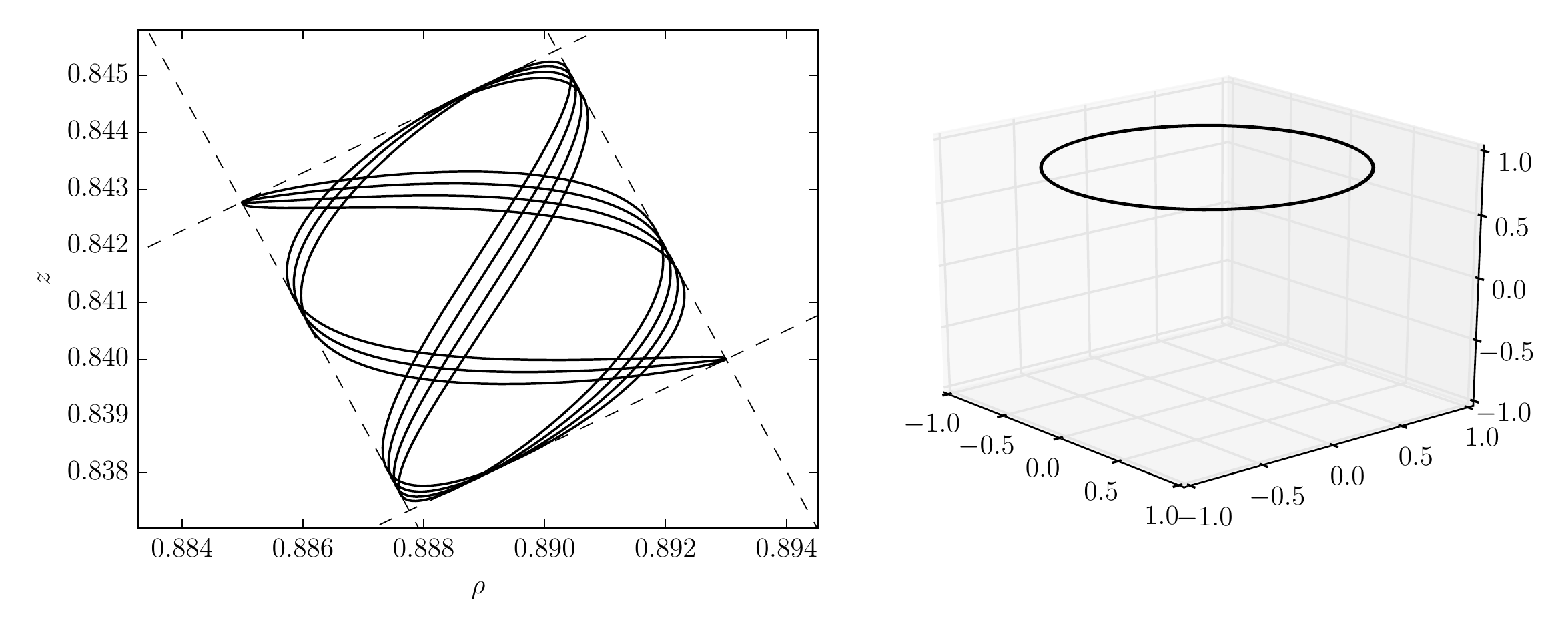}

\caption{Graphical depiction of a slightly perturbed displaced circular orbit in the E3BP far from the $xy$ plane. The initial conditions have been chosen close to the setup described by
eqs. \eqref{eq:d_z_bal} and \eqref{eq:v_d}. The left panel displays the trajectory in the $\left( \rho, z \right)$ plane (solid line),
together with the boundaries of the region of motion (dashed lines). The right panel displays the trajectory in the three-dimensional space.\label{fig:displaced}}
\end{figure*}

From the point of view of our solution to the E3BP, a displaced circular orbit is characterised by the solutions for $\xi$ and $\eta$ collapsing to constant functions.
According to eqs. \eqref{eq:xi_sol} and \eqref{eq:eta_sol}, this can happen only when
\begin{align}
f_\xi^\prime\left( \xi_r \right) & = 0,\\
f_\eta^\prime\left( \eta_r \right) & = 0.
\end{align}
From the physical point of view, the balance of the $z$ components of the two forces acting on the test particle leads to the following relation between $\rho$ and $z$:
\begin{equation}
\rho = \sqrt{\frac{\left(a+z\right)^2\left[\mu_1\left(a-z\right)\right]^\frac{2}{3}-\left(a-z\right)^2\left[\mu_2\left(a+z\right)\right]^\frac{2}{3}}
{\left[\mu_1\left(a-z\right)\right]^\frac{2}{3}-\left[\mu_2\left(a+z\right)\right]^\frac{2}{3}}}.\label{eq:d_z_bal}
\end{equation}
Given a pair of coordinates $\left(\rho_d,z_d\right)$ satisfying eq. \eqref{eq:d_z_bal}, we can then compute the magnitude of the total net force acting on the test particle
(parallel to the $xy$ plane and directed towards the $z$ axis) as
\begin{equation}
F_d = \frac{\mu_1\rho_d}{\left[\rho_d^2+\left(z_d-a\right)^2\right]^\frac{3}{2}} + \frac{\mu_2\rho_d}{\left[\rho_d^2+\left(z_d+a\right)^2\right]^\frac{3}{2}},
\end{equation}
$F_d$ corresponds to the force exerted by a virtual centre of attraction, lying at the coordinates $\left(0,z_d\right)$ in the $\left(\rho,z\right)$ plane, whose gravitational parameter
is
\begin{equation}
\mu_d = \rho_d^2F_d.
\end{equation}
We can now use the well-known relation between gravitational parameter and orbital radius for circular Keplerian orbits to compute the velocity along a circular displaced orbit
as
\begin{equation}
v_d = \sqrt{\frac{\mu_d}{\rho_d}}.\label{eq:v_d}
\end{equation}
Figures \ref{fig:spring} and \ref{fig:displaced} depict two representative slightly perturbed displaced circular orbits.

\section{Conclusions and future work}
\label{sec:conclusions}
In this paper we have presented for the first time a complete, closed-form solution to the three-dimensional problem of two fixed centres. Our solution is based
on the theory of Weierstrass elliptic and related functions, and it is expressed via unique formul\ae{} valid for any set of initial conditions
and physical parameters of the system. Remarkably, our solution is
strikingly similar to (albeit more complicated than) the solution of the two-body problem: the real time is substituted by a fictitious time, analogue to the mean
anomaly in Kepler's problem, and the connection between real and fictitious time is established by an equation structurally similar to Kepler's equation.

The compact form of our solution allows us to investigate the properties of the dynamical system. In particular, we have formulated analytical criteria
for quasi-periodic and periodic motion, and we have identified, via a simple numerical search, a few concrete representative quasi-periodic and periodic orbits.
We have also discussed the dichotomy between bounded and unbounded orbits, the topology of the regions of motion, and
we have identified displaced circular orbits as equilibrium points in elliptic-cylindrical coordinates.

In future papers, we will focus on the physical and astronomical applications of the results presented here. Of particular interest are the application of our
solution to the Vinti potential and the interpretation of the E3BP as a limiting case of the circular restricted three-body problem.

\bibliographystyle{mn2e}
\bibliography{biblio}

\begin{thebibliography}{}

\bibitem[\protect\citeauthoryear{Abramowitz \& Stegun}{Abramowitz \&
  Stegun}{1964}]{abramowitz_handbook_1964}
Abramowitz M.,  Stegun I.~A.,  1964, Handbook of mathematical functions with
  formulas, graphs, and mathematical tables.
Courier Dover Publications

\bibitem[\protect\citeauthoryear{Akhiezer}{Akhiezer}{1990}]{akhiezer_elements_1990}
Akhiezer N.~I.,  1990, Elements of the {Theory} of {Elliptic} {Functions}.
American Mathematical Society, Providence, RI

\bibitem[\protect\citeauthoryear{Aksenov, Grebenikov \& Demin}{Aksenov
  et~al.}{1962}]{aksenov_general_1962}
Aksenov Y.~P.,  Grebenikov Y.~A.,    Demin V.~G.,  1962, Planetary and Space
  Science, 9, 491

\bibitem[\protect\citeauthoryear{Aksenov, Grebenikov \& Demin}{Aksenov
  et~al.}{1963}]{aksenov_generalized_1963}
Aksenov Y.~P.,  Grebenikov Y.~A.,    Demin V.~G.,  1963, Soviet Astronomy, 7,
  276

\bibitem[\protect\citeauthoryear{Alfriend, Dasenbrock, Pickard \&
  Deprit}{Alfriend et~al.}{1977}]{alfriend_extended_1977}
Alfriend K.~T.,  Dasenbrock R.,  Pickard H.,    Deprit A.,  1977, Celestial
  mechanics, 16, 441

\bibitem[\protect\citeauthoryear{Arnold}{Arnold}{1989}]{arnold_mathematical_1989}
Arnold V.~I.,  1989, {Mathematical Methods of Classical Mechanics}, 2nd edn.
Springer

\bibitem[\protect\citeauthoryear{Besicovitch}{Besicovitch}{1932}]{besicovitch_almost_1932}
Besicovitch A.~S.,  1932, Almost {Periodic} {Functions}.
Cambridge University Press

\bibitem[\protect\citeauthoryear{Biscani \& Izzo}{Biscani \&
  Izzo}{2014}]{biscani_stark_2014}
Biscani F.,  Izzo D.,  2014, Monthly Notices of the Royal Astronomical Society,
  439, 810

\bibitem[\protect\citeauthoryear{Carinena, Ibort \& Lacomba}{Carinena
  et~al.}{1988}]{carinena_time_1988}
Carinena J.~F.,  Ibort L.~A.,    Lacomba E.~A.,  1988, Celestial Mechanics, 42,
  201

\bibitem[\protect\citeauthoryear{Charlier}{Charlier}{1902}]{charlier_mechanik_1902}
Charlier C. V.~L.,  1902, Die {Mechanik} des {Himmels}.
Leipzig, Veit

\bibitem[\protect\citeauthoryear{Coelho \& Herdeiro}{Coelho \&
  Herdeiro}{2009}]{coelho_relativistic_2009}
Coelho F.~S.,  Herdeiro C. A.~R.,  2009, Physical Review D, 80, 104036

\bibitem[\protect\citeauthoryear{Cordani}{Cordani}{2003}]{cordani_kepler_2003}
Cordani B.,  2003, The {Kepler} problem: group theoretical aspects,
  regularization and quantization, with application to the study of
  perturbations.
Birkhäuser Verlag, Basel; Boston

\bibitem[\protect\citeauthoryear{Darboux}{Darboux}{1901}]{darboux_sur_1901}
Darboux G.,  1901, Archives N\'{e}erlandaises des Sciences Exactes et
  Naturelles, 6, 371

\bibitem[\protect\citeauthoryear{Demin}{Demin}{1961}]{demin_orbits_1961}
Demin V.~G.,  1961, Soviet Astronomy, 4, 1005

\bibitem[\protect\citeauthoryear{Deprit}{Deprit}{1962}]{deprit_probleme_1962}
Deprit A.,  1962, Bull. Soc. Math. Belg., 14, 12

\bibitem[\protect\citeauthoryear{Gradshte\u{\i}n \& Ryzhik}{Gradshte\u{\i}n \&
  Ryzhik}{2007}]{gradshtein_table_2007}
Gradshte\u{\i}n I.~S.,  Ryzhik I.~M.,  2007, Table of Integrals, Series, And
  Products.
Academic Press

\bibitem[\protect\citeauthoryear{Greenhill}{Greenhill}{1959}]{greenhill_applications_1959}
Greenhill G.,  1959, The applications of elliptic functions.
Dover Publications

\bibitem[\protect\citeauthoryear{Halphen}{Halphen}{1886}]{halphen_traite_1886}
Halphen G.~H.,  1886, Trait\'{e} des fonctions elliptiques et de leurs
  applications.
Vol.~1, Paris, Gauthier-Villars

\bibitem[\protect\citeauthoryear{Hiltebeitel}{Hiltebeitel}{1911}]{hiltebeitel_problem_1911}
Hiltebeitel A.~M.,  1911, American Journal of Mathematics, 33, 337

\bibitem[\protect\citeauthoryear{Jones, Oliphant, Peterson et~al.,}{Jones
  et~al.}{2015}]{scipy}
Jones E.,  Oliphant T.,  Peterson P.,    et~al., 2001--2015, {SciPy}: Open
  source scientific tools for {Python}

\bibitem[\protect\citeauthoryear{Kraft}{Kraft}{1994}]{kraft_algorithm_1994}
Kraft D.,  1994, ACM Trans. Math. Softw., 20, 262

\bibitem[\protect\citeauthoryear{Lantoine \& Russell}{Lantoine \&
  Russell}{2011}]{lantoine_complete_2011}
Lantoine G.,  Russell R.,  2011, Celestial Mechanics and Dynamical Astronomy,
  109, 333

\bibitem[\protect\citeauthoryear{Namouni \& Guzzo}{Namouni \&
  Guzzo}{2007}]{namouni_accelerated_2007}
Namouni F.,  Guzzo M.,  2007, Celestial Mechanics and Dynamical Astronomy, 99,
  31

\bibitem[\protect\citeauthoryear{\'{O}'Math\'{u}na}{\'{O}'Math\'{u}na}{2008}]{omathuna_integrable_2008}
\'{O}'Math\'{u}na D.,  2008, Integrable {Systems} in {Celestial} {Mechanics}.
Springer Science \& Business Media

\bibitem[\protect\citeauthoryear{Pauli}{Pauli}{1922}]{pauli_uber_1922}
Pauli W.,  1922, Annalen der Physik, 373, 177

\bibitem[\protect\citeauthoryear{Saha}{Saha}{2009}]{saha_interpreting_2009}
Saha P.,  2009, Monthly Notices of the Royal Astronomical Society, 400,
  228–231

\bibitem[\protect\citeauthoryear{Siegel \& Moser}{Siegel \&
  Moser}{1971}]{siegel_lectures_1971}
Siegel C.~L.,  Moser J.~K.,  1971, {Lectures on Celestial Mechanics}.
Springer

\bibitem[\protect\citeauthoryear{Sundman}{Sundman}{1912}]{sundman_memoire_1912}
Sundman K.~F.,  1912, Acta Mathematica, 36, 105

\bibitem[\protect\citeauthoryear{Tannery \& Molk}{Tannery \&
  Molk}{1893}]{jules_tannery_elements_1893}
Tannery J.,  Molk J.,  1893, {\'{E}l\'{e}ments de la Theori\'{e} des Fonctions
  Elliptiques}.
Vol.~4, Gauthier Villars Et Fils Imprimeurs, Paris

\bibitem[\protect\citeauthoryear{Varvoglis, Vozikis \& Wodnar}{Varvoglis
  et~al.}{2004}]{varvoglis_two_2004}
Varvoglis H.,  Vozikis C.,    Wodnar K.,  2004, Celestial Mechanics and
  Dynamical Astronomy, 89, 343

\bibitem[\protect\citeauthoryear{Vinti}{Vinti}{1959}]{vinti_new_1959}
Vinti J.~P.,  1959, Journal of Research of the National Bureau of Standards,
  Section B: Mathematics and Mathematical Physics, 62B, 105

\bibitem[\protect\citeauthoryear{Waalkens, Dullin \& Richter}{Waalkens
  et~al.}{2004}]{waalkens_problem_2004}
Waalkens H.,  Dullin H.~R.,    Richter P.~H.,  2004, Physica D: Nonlinear
  Phenomena, 196, 265

\bibitem[\protect\citeauthoryear{Whittaker \& Watson}{Whittaker \&
  Watson}{1927}]{whittaker_course_1927}
Whittaker E.~T.,  Watson G.~N.,  1927, A Course of Modern Analysis, 4 edn.
Cambridge University Press

\end{thebibliography}

\appendix

\section{Solution algorithm}
In this section, we are going to detail the steps of a possible implementation of our solution to the three-dimensional E3BP, starting from
initial conditions in cartesian coordinates. The
algorithm outlined below requires the availability of implementations of the Weierstrassian functions $\wp$, $\wp^\prime$, $\wp^{-1}$,
$\zeta$ and $\sigma$, and of a few related ancillary functions (e.g., for the conversion of the invariants $g_2$ and $g_3$ to the periods).

The algorithm is given as follows:
\begin{enumerate}
 \item transform the initial cartesian coordinates into initial elliptic-cylindrical coordinates $\xi$, $\eta$ and $z$, and
 compute the initial Hamiltonian momenta $p_\xi$, $p_\eta$ and $p_\phi$. The formul\ae{} for these transformations are available
 in Section \ref{sec:formulation};
 \item compute the constants of motion $h$, $h_\xi$ and $h_\eta$, through the substitution of the initial Hamiltonian
coordinates and momenta into eqs. \eqref{eq:t_Ham}, \eqref{eq:h_xi} and \eqref{eq:h_eta};
 \item compute the values of the invariants $g_2$ and $g_3$ for both $\wp_\xi$ and $\wp_\eta$, using eqs. \eqref{eq:g2_xi}, \eqref{eq:g3_xi},
 \eqref{eq:g2_eta} and \eqref{eq:g3_eta}. The periods of $\wp_\xi$ and $\wp_\eta$ can be determined from the invariants using known formul\ae{}
 \citep[see, e.g.,][Chapter 18]{abramowitz_handbook_1964};
 \item determine the roots $\xi_r$ and $\eta_r$ of the polynomials \eqref{eq:xi_poly} and \eqref{eq:eta_poly}. Since these are quartic polynomials,
 there are 4 possible values for each root. We can immediately discard complex roots and those roots whose values are outside the domains of interest
 for the variables (i.e., $\left(1,+\infty\right)$ for $\xi$ and $\left(-1,1\right)$ for $\eta$). Note that complex roots and roots outside the domain
 of physical interest can still be used for the computation of the solution, but they lead to complex-valued times of pericentre passage and they have no immediate
 physical interpretation;
 \item using the surviving values for $\xi_r$ and $\eta_r$, compute the times of pericentre passage $\tau_\xi$ and $\tau_\eta$ via eqs. \eqref{eq:tau_xi_0} and \eqref{eq:tau_eta_0}.
 As explained at the very end of Subsection \ref{subsec:xi_sol}, the calculation of $\tau_\xi$ and $\tau_\eta$ via $\wp^{-1}$ produces a pair of values for each $\xi_r$ 
 and $\eta_r$. We can immediately discard the values of $\xi_r$ and $\eta_r$ which result in complex $\tau_\xi$ and $\tau_\eta$, for they correspond to unreachable roots;
 \item at this point either one (for unbounded motion) or two (for bounded motion) values for $\xi_r$ and $\eta_r$ remain, both real. We select the smaller values for $\xi_r$ and $\eta_r$ (as we are
 interested in the times of \emph{pericentre} passage), and we compute the corresponding pairs of values for $\tau_\xi$ and $\tau_\eta$ via eqs. \eqref{eq:tau_xi_0} and \eqref{eq:tau_eta_0}.
 These pairs of values will all be real and positive. We select the smaller values for both times of pericentre passage and we check the sign of the initial values for $d\xi/d\tau$
 and $d\eta/d\tau$: if they are positive, we negate the corresponding time of pericentre passage (as the pericentre was reached \emph{before} $\tau=0$), otherwise we can leave them
 as they are (as the pericentre will be reached \emph{after} $\tau=0$);
 \item after having determined $\xi_r$, $\tau_\xi$, $\eta_r$ and $\tau_\eta$ we have all the ingredients to implement the solutions for the three coordinates, their conjugate momenta
 and the time equation (eqs. \eqref{eq:xi_sol}, \eqref{eq:eta_sol}, \eqref{eq:phi_tau_sol}, \eqref{eq:t_sol}, \eqref{eq:pxi_sol} and \eqref{eq:peta_sol}).
\end{enumerate}
The procedure outlined above assumes the following conventions:
\begin{itemize}
 \item the period pairs for $\wp_\xi$ and $\wp_\eta$ are chosen as explained in Subsection \ref{subsec:xi_sol}: the first period is always real and positive, the second one is always complex
 with positive imaginary part;
 \item the $\wp^{-1}$ function returns a pair of values within the fundamental parallelogram defined by the periods pair.
\end{itemize}
\section{Code availability}
We have implemented our solution to the E3BP in an open-source Python module which is freely available for download here:
\newline
\newline
\url{https://github.com/bluescarni/e3bp}
\newline
\newline
The module depends on the \texttt{w\_elliptic} Python/C++ library, which provides an implementation of the Weierstrassian functions. The code is available here:
\newline
\newline
\url{https://github.com/bluescarni/w\_elliptic}
\newline
\newline
All the numerical computations and all the graphs presented in the paper have been implemented with and produced by these two software modules.

\bsp

\label{lastpage}

\end{document}